\documentclass[12pt]{elsarticle}  
\usepackage{epsfig,xfrac,booktabs,xfrac,amsmath,graphicx}
\usepackage{caption}
\usepackage{subcaption}
\usepackage{multicol}
\usepackage{multirow}

\newcommand\sr{\mathrm{sr}}
\newcommand\ini{\mathrm{ini}}
\newcommand\en{\mathrm{end}}

\newcommand\sca{\mathrm{S}}
\newcommand\ten{\mathrm{T}}
\newcommand\Pl{\mathrm{Pl}}

\newcommand\D{\mathrm{d}}

\newcommand\e{\mathrm{e}}

\begin{document}

\title{Some inflationary models  under the light of  Planck 2018 results}

\author{Daniel Pozo$^1$, Jordan Zambrano$^1$,  Ismael Villegas$^1$,  Rafael Hern\'{a}ndez--Jim\'enez$^2$, and Clara Rojas$^1$}

\address{$^1$ Yachay Tech University, School of Physical Sciences and Nanotechnology, Hda. San Jos\'e s/n y Proyecto Yachay, 100119, Urcuqu\'i, Ecuador}
\ead{crojas@yachaytech.edu.ec}

\address{$^2$ Departamento de F\'isica,
Centro Universitario de Ciencias Exactas e Ingenier\'ia, Universidad de Guadalajara\\
Av. Revoluci\'on 1500, Colonia Ol\'impica C.P. 44430, Guadalajara, Jalisco, M\'exico}

\begin{abstract}

In this work we study four well--known inflationary scenarios that are reported by the most recent Planck observations: Natural inflation, Hilltop quartic inflation, Starobinsky inflationary model, and Large field power--law potentials $V(\phi)\sim \phi^{p}$, considering $p=\sfrac{2}{3}, \sfrac{4}{3}$. The analysis is done using both the slow--roll approximation and the numerical solution to the background and perturbation equations. We show that the numerical solution improved the precision of these models with respect to the contour plot $r$ vs. $n_\sca$, having a lower $r$ in each model compared to the value calculated from the slow--roll approximation. \\

\end{abstract}

\maketitle

\section{Introduction}

Inflation is an epoch of accelerated expansion of the Universe introduced in the $1980$'s to solve the shortcomings of the Big Bang \cite{Albrecht:1982wi,guth:1981, Linde:1981mu,Starobinsky:1980te, Sato:1980yn}. The most important of these issues are: the horizon problem, the flatness problem, and the unwanted magnetic monopoles. The first one related the to the fact of the limit of the speed of light and treats with the causality of homogeneous distribution of heterogeneity of the temperature seen in the CMB image, the second one treats a fine--tuning problem about the curvature of the space at very few moments from the big bang \cite{weinberg:2008}, and the last one arises from the fact that the expected density of magnetic monopoles from calculations of Grand Unified Theory (GTU) does not match with the never recorded measurement of such relics \cite{vazquez:2020,preskill:1979}. 

These fine tuning inconveniences not only were precursors of doubts and incompleteness regarding the hot big bang theory, but also create some philosophical inquiries about how these so unlike circumstances happened to give place to our universe. In that way, inflation plays a crucial role not only in giving an explanation for these conditions but making them into consequences.

Moreover, inflation also presents two relevant features, provided its quantum fluctuations. They yield a spectrum of scalar perturbations and a primordial background of gravitational waves. Therefore, we have a quantum perturbation theory of the isotropic and homogeneous Friedmann--Lema\^{i}tre--Robertson--
Walker (FLRW) universe \cite{odintsov:2023,martin:2008}. By studying these scalar and tensor perturbations, we can model and contrast them with observational Planck data~\cite{akrami:2020}. Remarkably, these quantum fluctuations could be the seeds of the initial conditions observed throughout the large--scale structure of the universe~\cite{habib:2005b}. Therefore, understanding their spectrum allows us to select models that produce the best fit with respect to observational data. Recently, a numerical method based in a Mote Carlo approach has been proposed in the literature to give theoretical predictions of the observables such as the amplitude of the scalar and tensor perturbations and the spectral index for scalar perturbations \cite{giare:2023c}.

\begin{figure}
\includegraphics[scale=0.8]{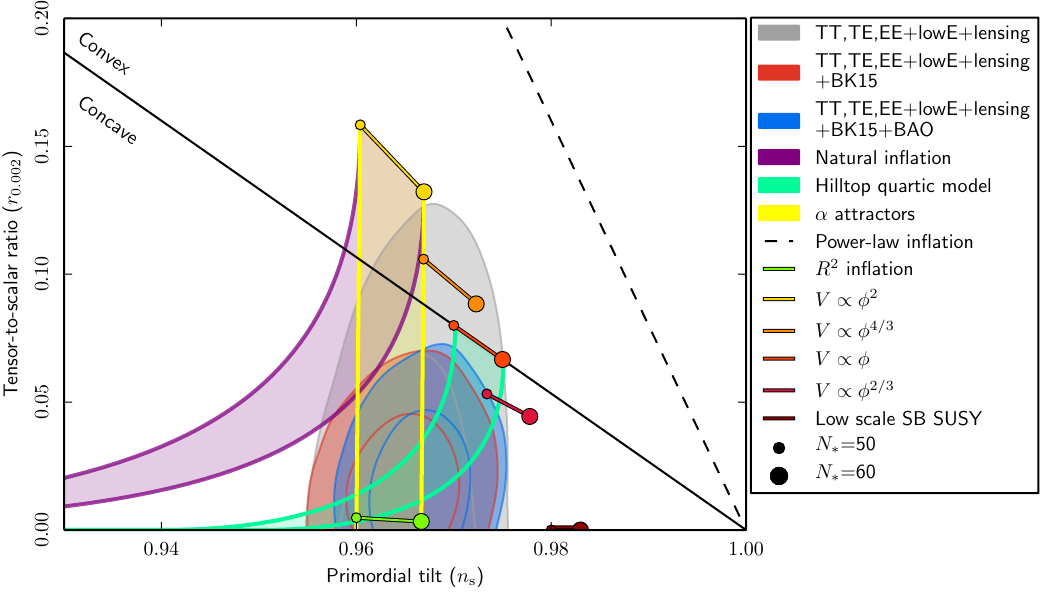}
\caption{Contour ($n_\sca,r)$ showing the Planck $2018$ data and the slow--roll calculations for some inflationary models. Figure taken from Ref. \cite{akrami:2020}.}
\label{fig:Planck}
\end{figure}

Throughout the years, several single field inflation scenarios have been proposed in the literature \cite{german:2023,martin:2014,martin:2006}, but only a few are supported by the Planck $2018$ results \cite{akrami:2020}. All are represented by the potential of the inflaton ($\phi$) $V=V(\phi)$. The upshots of such survey allow us to discern among them; in fact, there is evidence that Planck's outcome favors concave inflationary potentials, as we can observe in Fig. \ref{fig:Planck}, where it is shown the primordial tilt $n_\sca$ and tensor--to--scalar tensor $r$, as well as their correspondence levels of confidence contour plots, taken from Planck $2018$ data~\cite{akrami:2020}. 

In this work, we study four inflationary models already analyzed by Planck $2018$ \cite{akrami:2020}: Natural inflation \cite{freese:1990, adams:1993}; Hilltop inflation \cite{boubekeur:2005}; Starobinsky inflationary model \cite{starobinsky:1980}; and Large--Field inflation \cite{linde:1983}. Note that two of these scenarios are well inside the Planck contours: Hilltop quartic and Starobinsky (or $R^2$). And here $n_\sca=0.9649 \pm 0.0044$ and small values for $r$. 

According to background dynamics, accelerated expansion occurs when the potential energy of the field, $V(\phi)$, dominates over its kinetic counterpart, $\dot{\phi}^{2}/2$. This approach is called slow--roll inflation. We proceed using this method to obtain the initial and final states of the inflaton $\phi$ values, taking into account the cases for $N = 50$, and $N = 60$. Once we have obtained such estimates, we use them as initial conditions for our numerical implementation.   

This article is structured as follows: in Section \ref{equations} we introduce the background and perturbations equations. Then in Section \ref{slow-roll} we present the slow--roll analysis and the formulas of the observables under this approximation. In Section \ref{numerical} we follow our numerical procedure to solve the aforementioned system of equations. Next, in Section \ref{models} we show the corresponding analysis of each inflationary model considered in this project. Therefore, in Section \ref{results} we examine and discuss our results. Finally, in Section \ref{conclusions} we present our conclusions of this work.

\section{Equations of motion of Background and Perturbations}
\label{equations}

The equations of motion of the inflaton $\phi$ and the Hubble parameter $H$ (with $M_\textnormal{Pl}=1$) are given by \cite{liddle:2000},

\begin{eqnarray}
\label{H^2}
H^2&=&\frac{1}{3}\left[V(\phi)+\dfrac{1}{2}\dot{\phi}^2\right],\\
\label{ddotphi}
\ddot{\phi}+3H\dot{\phi}&=&-\dfrac{\partial V(\phi)}{\partial\phi},
\end{eqnarray}

where the dots indicate derivatives with respect to physical time $t$, and $\partial V(\phi)/\partial\phi$ is the derivative of the potential energy with respect to the $\phi$.
 
Eq. \eqref{H^2} is the Friedmann equation in terms of the scalar field, and Eq. \eqref{ddotphi} represents the inflaton dynamics.

The scalar perturbations formulas come from a quantum perturbation theory, as well as linear fluctuations to the FLRW metric; having~\cite{habib:2005b},

\begin{equation}
\label{dotdotuk}
u_k''+\left(k^2-\dfrac{z_{S}''}{z_{S}}\right)u_k=0,
\end{equation}

where $z_{S}=a\phi'/\mathcal{H}$, $\mathcal{H}=a'/a$. Here, the prime indicates derivative with respect to the conformal time $\tau$. The relation between $t$ and $\tau$ is given by equation $\D t=a\,\D \tau$.

For tensor perturbations, one introduces the function $v_k=ah$, where $h$ represents the amplitude of the gravitational wave. Tensor perturbations obey a second order differential equation given by~\cite{habib:2005b},

\begin{equation}
\label{dotdotvk}
v_k''+\left(k^2-\dfrac{a''}{a}\right)v_k=0.
\end{equation}

Considering the limits $k^2\gg|z_{S}''/z_{S}|$ (short wavelength) and $k^2\ll|z_{S}''/z_{S}|$  (long wavelength), we have the solutions of Eq. (\ref{dotdotuk}) present the following asymptotic behavior,

\begin{equation}
\label{boundary_0}
u_k\rightarrow \dfrac{e^{-ik\tau}}{\sqrt{2k}}
\quad \left(k^2\gg|z_{S}''/z_{S}|, -k\tau\rightarrow \infty \right),
\end{equation}

\begin{equation}
\label{boundary_i} u_k\rightarrow A_k z  \quad \left(k^2\ll|z_{S}''/z_{S}|,-k\tau\rightarrow 0\right).
\end{equation}

\noindent And the same asymptotic conditions hold for tensor
perturbations. Hereafter Eq. \eqref{boundary_0} will be used as the initial condition for the perturbations. 

On the other hand, the power spectra of the scalar and tensor perturbations are given by the expressions,

\begin{eqnarray}
\label{PS}
P_\sca(k)&=& \lim_{k\tau\rightarrow 0^{-}} \dfrac{k^3}{2 \pi^2}\left|\dfrac{u_k(\tau)}{z_{S}(\tau)} \right|^2,\\
\label{PT}
P_\ten(k)&=& \lim_{k\tau\rightarrow 0^{-}} \dfrac{k^3}{2 \pi^2}\left|\dfrac{v_k(\tau)}{a(\tau)} \right|^2,
\end{eqnarray}

and the scalar spectral index together with its running are defined by,

\begin{eqnarray}
\label{nS}
n_\sca(k) &=& 1+\dfrac{\D\ln P_\sca(k)}{\D\ln k},\\
\alpha_\sca(k) &=&\dfrac{\D \,n_\sca(k)}{\D \ln k}.
\end{eqnarray}

In addition, the tensor--to--scalar ratio $r$ is defined as,

\begin{equation}
\label{r}
r=8\dfrac{P_\ten(k)}{P_\sca(k)}.
\end{equation}

Notice also that for single field models, the tensor--to--scalar ratio $r$ can be related to the tilt of the tensor spectrum of perturbations, this is the tensor spectral index $n_\ten$, by the consistency equation \cite{copeland:1993},

\begin{equation}
\label{nT}
n_\ten(k)=-\dfrac{r}{8}.
\end{equation}

\section{Slow--roll analysis}
\label{slow-roll}

The slow--roll approximation is considered the standard technique used in inflation. This approach considers that $V(\phi) \gg \dot{\phi}$; additionally by differentiating this suggests the further condition $\mid\ddot{\phi}\mid\ll\mid V_{\phi}\mid$; however, note that derivatives of approximations need not themselves be valid approximations, and so this is an additional condition. Then we also neglect the $\ddot{\phi}$ term from Eq. \eqref{ddotphi}. Thus, the background equations of motion \eqref{H^2} and \eqref{ddotphi} are simplified in the following way,

\begin{eqnarray}
\label{Hsr}
H^2&\simeq&\dfrac{1}{3}V(\phi),\\
\label{phisr}
3H\dot{\phi}&\simeq&-\dfrac{\partial V(\phi)}{\partial\phi}.
\end{eqnarray}

An inflationary model is sometimes parameterized by expanding the potential in a Taylor series, i.e., in higher and higher derivatives of $V(\phi)$. The first two so called slow--roll parameters,

\begin{eqnarray}
\label{epsilon}
\epsilon&=&\dfrac{1}{2}\left(\dfrac{V'}{V}\right)^2,\\
\label{eta}
\eta &=& \dfrac{V''}{V},
\end{eqnarray}
where the prime indicates derivative respect to $\phi$. The first one measures the slope of the potential, and the second one the curvature. The amount by which the universe inflates is measured as the number of e--foldings, giving by,

\begin{equation}
\label{N}
N \simeq \int_{\phi_{\textnormal{end}}}^\phi \dfrac{V}{V'} \D \phi,
\end{equation}
where $\phi_\textnormal{end}$ is the value of the scalar field at the end of inflation.

Moreover, the scalar power spectrum $P_\sca(k)$ and the tensor power spectrum $P_\ten(k)$ from the slow--roll approximation are given by the expressions \cite{adshead:2011},

\begin{eqnarray}
\label{sr_PS}
P_\sca(k)&\simeq&\dfrac{1}{12\pi^2}\dfrac{V^3}{V'^2}
\left[1-\left(2 C+\dfrac{1}{6}\right)\epsilon + \left( C-\dfrac{1}{3}\right)\eta\right],\\
\label{sr_PT}
P_\ten(k)&\simeq&\left[ 1-\left(C+1\right)\epsilon_H\right]^2\left(\dfrac{H}{2\pi}\right)^2\bigg. \bigg |_{k=aH},
\end{eqnarray}
where $C=-2+\ln 2 + b\simeq- 0.729637$, here $b$ is the Euler--Mascheroni constant, and

\begin{equation}
\epsilon_H=\epsilon\left(1-\dfrac{2}{3}\epsilon+\dfrac{1}{3}\eta\right).
\end{equation}

Furthermore, the scalar spectral index $n_\sca(k)$ and the tensor--to--scalar ratio $r$ from the slow--roll approximation are given, in terms of the slow--roll parameters, by \cite{liddle:2000},

\begin{eqnarray}
\label{sr_nS}
n_\sca&\simeq&1-6\epsilon+2\eta,\\
\label{sr_r}
r&\simeq& 16\epsilon.
\end{eqnarray}

Expressions \eqref{sr_PS}, \eqref{sr_PT}, \eqref{sr_nS}, and \eqref{sr_r} must be to evaluate at horizon crossing $k=aH$, and explicitly depend on the inflaton value at that instance $\phi_{*}$, which depends on time $t_{*}$. Thus, for each $k$ one obtains a value of $\phi_{*}(t_{*})$ that is then substituted into Eqs. \eqref{sr_PS}, \eqref{sr_PT}, \eqref{sr_nS}, and \eqref{sr_r}.

Finally, from an additional approximation of the scalar power--spectrum $P_\sca(k)$, that is \cite{liddle:2000},

\begin{equation}
\label{M}
\delta_R=\dfrac{1}{24 \pi^2}\dfrac{V(\phi)}{\epsilon(\phi)},
\end{equation}
we calculate the free parameter $M$, that characterizes each inflationary model, by fixing $\delta_R \sim 2.1 \times 10^{-9}$ (COBE normalization), at the pivot scale $k_*=0.05$ Mpc$^{-1}$  \cite{ragavendra:2023}.

\section{Numerical analysis}
\label{numerical}
 
In this section, we present our numerical implementation. To solve the perturbation equations numerically, first we have to solved the equations of motion Eq. \eqref{H^2} and Eq. \eqref{ddotphi}  numerically with respect to physical time $t$ taking as initial conditions the slow--roll solutions. Hence, the scale factor $a_\sr(t)$, the scalar field $\phi_\sr(t)$, and the derivative with respect to the physical time of the scalar field $\phi_\sr$  are obtained into the slow--roll approximation, and evaluated at $t=0$. The integration is made from $t_\ini=0$ until the end of inflation $t_\en$, whic is calculated in each case. From the four model studied only three have analytical solution for the slow--roll approximation: the Natural, Starobinsky and large field inflationary models. In the case of the Hilltop potential we have to solve numerically the slow--roll equations Eqs. \eqref{Hsr} and \eqref{phisr}, in this case the integration is also made from $t_\ini=0$ until the end of inflation $t_\en$.

On the other hand, the equations of the linear fluctuations are $\tau$ dependent, so we proceed to rewrite them in terms of the variable $t$. Therefore, these equations are,

\begin{eqnarray}
\label{dotu}
\ddot{u_k}+\frac{\dot{a}}{a}\dot{u_k}+\frac{1}{a^2}\left[k^2-\frac{\left(\dot{a}\dot{z_\sca}+a\ddot{z_\sca}\right)a}{z_\sca} \right]u_k&=&0,\\
\label{dotv}
\ddot{v_k}+\frac{\dot{a}}{a}\dot{v_k}+\frac{1}{a^2}\left[k^2-\left(\dot{a}^2+a\ddot{a}\right) \right]v_k&=&0,
\end{eqnarray}
where $z_\sca$, $\dot{z}_\sca$, and $\ddot{z}_\sca$ are given in terms of $a$, $\phi$, and their time $t$ derivatives,

\begin{eqnarray}
\label{zs}
z_\sca&=&\dfrac{a^2 \dot{\phi}}{\dot{a}},\\
\dot{z}_\sca&=&a\dot{\phi}\left(2-\dfrac{a \ddot{a}}{\dot{a}^2} \right)+\dfrac{a^2\ddot{\phi}}{\dot{a}},\\
\nonumber
\ddot{z}_\sca &=& 2\dot{a}\dot{\phi}+\dfrac{2a^2\ddot{a}^2\dot{\phi}}{\dot{a}^3}+4a\ddot{\phi}-\dfrac{a^2\left(2\ddot{a}\ddot{\phi}+\dddot{a}\dot{\phi}\right)}{\dot{a}^2}\\
&& + \dfrac{a\left(-2\ddot{a}\dot{\phi}+a\dddot{\phi}\right)}{\dot{a}}.
\end{eqnarray}

The equations for the scalar and tensor perturbations \eqref{dotu} and \eqref{dotv}, respectively, are then numerically integrated. Note that $u_k$ and $v_k$ are complex functions, then two differential equations are solved for each of them, both real and imaginary parts. 
Integration is performed in two instances. The first part is done in the limit when $k^2\gg\sfrac{\left(\dot{a}\dot{z_\sca}+a\ddot{z_\sca}\right)a}{z_\sca} $, and $k^2\gg \dot{a}^2+a\ddot{a}$ for scalar and tensor perturbations, respectively, so the differential equation to solved are the following,

\begin{eqnarray}
\label{dotuk_k2}
\ddot{u_k}&+\dfrac{\dot{a}}{a}\dot{u_k}+\dfrac{k^2}{a^2} u_k=0,\\
\label{dotvk_k2}
\ddot{v_k}&+\dfrac{\dot{a}}{a}\dot{v_k}+\dfrac{k^2}{a^2} v_k=0,
\end{eqnarray}
the integration is done from $t_\ini=0.001\, h_c(k)$ until $t_\en=0.05\, h_c(k)$, where $h_c(k)$ is the horizon defined by the time where the mode freezes $k=a H$. In this stage, we have used as initial condition Eq. \eqref{boundary_0} but written in terms of $t$, that is $\tau\simeq -1/[a(t)H]$ since $H$ is nearly constant inside the horizon {\footnote{In particular, during inflation,

\begin{equation*}\label{conformal-time-inflation}
\tau=\int^{a}_{0}\frac{da}{Ha^{2}}\simeq \frac{1}{H}\int^{a}_{0}\frac{da}{a^{2}}\simeq -\frac{1}{aH}. 
\end{equation*}  
}}, then $u_k$ and $v_k$ exhibits an oscillatory behavior.

Then, we use the final stage of this solution as an initial condition to solve the complete set of Eqs. \eqref{dotu} and \eqref{dotv} from $k=0.0001$\,Mpc$^{-1}$ to $k=10$\,Mpc$^{-1}$, and the integration is done from $t_\ini=0.05\, h_c(k)$ until $t_\en=3\, h_c(k)$, when the perturbations are already frozen. Finally, from Eqs. \eqref{PS} and \eqref{PT} we  calculated the scalar and tensor power spectra.

Also, in order to find the scalar spectral index $n_\sca(k)$ we implement the fit of the scalar power spectrum $P_\sca(k)$ with a power--law form \cite{giare:2023b,vazquez:2013,das:2023}, that is,

\begin{equation}
\label{PS_fit}
P_\sca(k)=A_\sca\left(\dfrac{k}{k_*}\right)^{n_\sca-1+\frac{1}{2}\alpha_\sca\ln\left(\frac{k}{k_*}\right)},
\end{equation}
so that $P_\sca(k)$ becomes scale dependent. Here, $A_\sca$ is the amplitude of the scalar power spectrum, and $\alpha_\sca$ is the running of the scalar spectral index. Equation \eqref{PS_fit} is evaluated at a given pivot scale $k_*=0.002$\,Mpc$^{-1}$, and $k_*=0.05$\,Mpc$^{-1}$.

Then, to find the tensor--to--scalar ratio $r$ we also utilize the fit of the tensor power spectrum $P_\ten(k)$ with a power--law form \cite{vazquez:2020,finelli:2018,vazquez:2013},

\begin{equation}
\label{PT_fit}
P_\ten(k)= A_\ten\left(\dfrac{k}{k_*}\right)^{n_\ten+\frac{1}{2}\alpha_\ten\ln\left(\frac{k}{k_*}\right)},
\end{equation}
where $A_\ten$ is the amplitude of the tensor power spectrum, and $\alpha_\ten$ is the running of the tensor spectral index. Equation \eqref{PT_fit} is evaluated at a given pivot scale $k_*=0.002$\,Mpc$^{-1}$.

The relation between the scalar power spectrum $P_\sca(k)$ and the amplitude of the scalar power spectrum $A_\sca$, without considering the running of the scalar spectral index $\alpha_\sca$, is given by \cite{akrami:2020},

\begin{equation}
\label{AS}
\ln P_\sca(k)=\ln A_\sca+\left(n_\sca-1\right)\ln\left(\dfrac{k}{k_*}\right),
\end{equation}
here Planck takes as pivot scale $k_{*}=0.05$\,Mpc$^{-1}$ \cite{akrami:2020}. Finally, we evaluate the scalar power spectrum $P_\sca(k)$ at $k=0.05$\,Mpc$^{-1}$, therefore $P_\sca(k)$ becomes equal to $A_\sca$.

\section{Models of inflation}
\label{models}

\subsection{Natural Inflation}

Natural inflation was introduced in $1990$ \cite{freese:1990, adams:1993} as a proposal to generate an inflationary epoch. Although this model was consistent with Planck $2015$ data \cite{freese:2015}, currently it is disfavored by the newest version Planck $2018$ plus the BK$15$ data \cite{akrami:2020,stein:2022}. However, considering a prolonged reheating period, natural inflation returns to the $95\%$ of confidence levels of Planck observations \cite{nina:2022}. Furthermore, in the warm inflation scenario, natural inflation is consistent with the constraints of $n_\sca$ and $r$ from Planck $2018$ data \cite{montefalcone:2023}.

\begin{figure}[th!]
\centering
\includegraphics[scale=0.5]{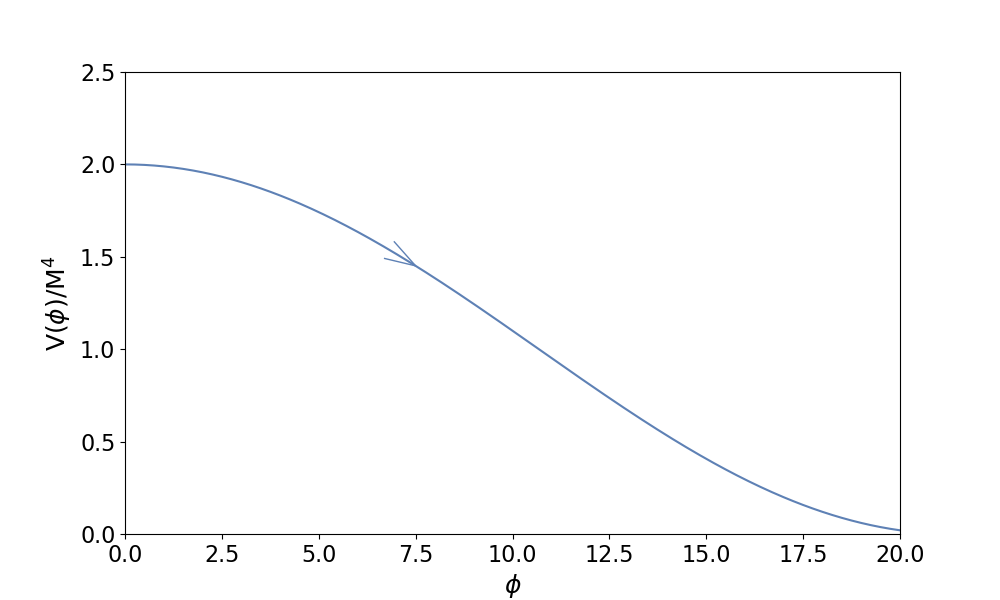}
\caption{Inflationary potential for Natural Inflation using $f=6.8$.}
\label{fig:natural_V}
\end{figure}

The natural inflation potential is given by \cite{cook:2023,akrami:2020,martin:2014},

\begin{equation}
\label{natural_V}
V(\phi)=M^4 \left[1+\cos\left(\dfrac{\phi}{f}\right)\right],
\end{equation}
where $M$ is a free parameter to be determined by the normalization of the scalar power spectrum $P_\sca(k)$, and $0.3 < \log_{10} f < 2.5$ \cite{akrami:2020}. The behavior of the potential is shown in Fig. \ref{fig:natural_V}. Inflation occurs from left to right, so the field $\phi$ slowly rolls near a maximum of the potential to its minimum.

From Eqs. \eqref{epsilon} and \eqref{eta} we calculate analytically the slow--roll parameters, yielding,

\begin{eqnarray}
\label{natural_epsilon}
\epsilon&=&\dfrac{1}{2f^2}\dfrac{\sin^2\left(\frac{\phi}{f}\right)}{\left[ 1+\cos\left(\frac{\phi}{f}\right)\right]^2},\\
\label{natural_eta}
\eta&=&-\dfrac{1}{f^2} \dfrac{\cos\left(\frac{\phi}{f}\right)}{\left[ 1+\cos\left(\frac{\phi}{f}\right)\right]}.
\end{eqnarray}

When $\epsilon=1$ inflation ends, so the value for the scalar field at the end of inflation is given by,

\begin{equation}
\label{natural_phiend}
\phi_\textnormal{end}=f \arccos\left(\dfrac{1-2f^2}{1+2f^2}\right). 
\end{equation}

The number of e--foldings is calculated from equation \eqref{N}, that is,

\begin{equation}
\label{natural_N}
N\simeq f^2 \ln\left[\dfrac{1-\cos\left(\sfrac{\phi_\textnormal{end}}{f}\right)}{1+\cos\left(\sfrac{\phi_\textnormal{end}}{f}\right)}\right]. 
\end{equation}

Fixing the number of e--foldings between $50-60$ e--folds we can obtain $\phi_{\textnormal{ini}}$ as,

\begin{equation}
\label{natural_phiini}
\phi_\textnormal{ini}= f \arccos\left\{1-\left[ 1-\cos\left(\frac{\phi_\textnormal{end}}{f}\right)\right]e^{-\sfrac{N}{f^2}}\right\}.
\end{equation}

Replacing $\phi_\textnormal{end}$ [from Eq. \eqref{natural_phiend}] in Eq.  \eqref{natural_phiini} we obtain,

\begin{equation}
\label{natural_phiini_def}
\phi_\textnormal{ini}= f \arccos\left(1- \dfrac{4f^2}{1+2f^2}e^{-\sfrac{N}{f^2}}\right).
\end{equation}

Then, using the slow--roll approximations, Eqs. \eqref{Hsr} and \eqref{phisr} the scale factor $a_{\sr}(t)$ and the scalar field $\phi_{\sr}(t)$ become,

\begin{eqnarray}
\label{natural_asr}
\nonumber
a_\sr(t)&\simeq& \exp\left\{-2f^2\ln\left\{\cosh\left[\textnormal{arctanh}\left(\dfrac{\sqrt{1+\cos\left(\sfrac{\phi_\textnormal{ini}}{f}\right)}}{\sqrt{2}}\right)-\dfrac{M^2}{\sqrt{6} f^2} t\right]\right\}\right.\\
&-&\ln\left\{\cosh\left[\textnormal{arctanh}\left(\dfrac{\sqrt{1+\cos\left(\sfrac{\phi_\textnormal{ini}}{f}\right)}}{\sqrt{2}}\right)\right\}\right\},\\
\nonumber
\label{natural_phisr}
\phi_\sr(t)&\simeq& f \arccos\left\{2\tanh\left[\textnormal{arctanh} \left(\dfrac{\sqrt{1+\cos\left(\sfrac{\phi_\textnormal{ini}}{f}\right)}}{\sqrt{2}}\right) -\dfrac{M^2}{\sqrt{6} f^2}t\right]^2-1\right\},\\
\end{eqnarray}
here $M$ is computed from Eq. \eqref{M}, yielding,

\begin{equation}
\label{natural_M}
M^4=\dfrac{12 \delta_R \pi^2 \sin^2\left(\frac{\phi_*}{f}\right)}{f^2 \left[1+\cos\left(\frac{\phi_*}{f}\right)\right]^3},
\end{equation}
where $\phi_*$ is calculated at the horizon crossing $k=0.05$ Mpc$^{-1}$.

To solved the equation of motion numerically we integrated Eqs. \eqref{H^2} and \eqref{ddotphi}, using as initial condition the derivative of Eq. \eqref{natural_phisr} evaluated at $t=0$, $\dot\phi_{\sr}(0)$. The integration was done since $t_{\ini}=0$ until $t_{\en}$. The initial condition for $\dot\phi_{\sr}(0)$ is given by,

\begin{equation}
\dot\phi_{\sr}(0)=-\dfrac{M^2\left[-1+\cos\left(\frac{\phi_\ini}{f}\right)\right] \sqrt{1+\cos\left(\frac{\phi_\ini}{f}\right)}}{\sqrt{3} f \sin\left(\frac{\phi_\ini}{f}\right)}.
\end{equation}

In table \ref{table:natural_tend} we show the value of $t_{\en}$ for each value of $f$ that we use in our contour plots for the Natural inflation model $N=50$, and $N=60$. 

\begin{table}[th!]
\begin{center}
\begin{tabular}{c | c | c }
\toprule 
N & $f$ & $t_\en \times 10^6$ $\left[\textnormal{M}_\Pl ^{-1}\right]$\\
\midrule 
\multirow{8}{*}{$50$}  & 4 & $3.34115$  \\
                       & 5 & $2.61706$ \\
                       & 6 & $2.34349$ \\
                       & 7 & $2.21186$ \\
                       & 9 & $2.09330$ \\
                       & 12&  $2.02757$\\
                       & 20&  $1.98117$\\
                       & 100& $1.95885$\\
\midrule
\multirow{8}{*}{$60$}  & 4 & $5.37998$\\
                       & 5 & $3.87416$ \\
                       & 6 & $3.34232$ \\
                       & 7 & $3.09656$ \\
                       & 9 & $2.88293$ \\
                       & 12&  $2.76873$\\
                       & 20&  $2.69075$\\
                       & 100& $2.65349$\\
\bottomrule
\end{tabular}
\caption{Value of the time at the end of inflation $t_\en$ for each value of $f$ in the Natural inflation scenario.}
\label{table:natural_tend}
\end{center}
\end{table}

\newpage
The scalar power spectrum $P_\sca(k)$, the scalar spectral index $n_\sca(k)$, and the tensor--to--scalar ratio $r$ are calculated from Eqs. \eqref{sr_PS}, \eqref{sr_nS}, and \eqref{sr_r}; and they are,

\begin{eqnarray}
\label{natural_PSsr}
\nonumber
P_\sca(k)&=&\dfrac{M^{4}}{144  \pi^2} \left[-1-12C+12 f^2+ \left(5+12 f^2\right)\cos\left(\frac{\phi_*}{f}\right)\right]\cot^2\left(\frac{\phi_*}{2f}\right) \,,\\\\
\label{natural_nS}
n_\sca(k)&=&\dfrac{1}{f^2}\left[1+f^2-2\sec^2\left(\frac{\phi_*}{2f}\right)\right],\\
\label{natural_r}
r&=&\dfrac{8}{f^2}\tan^2\left(\dfrac{\phi_*}{2f}\right),
\end{eqnarray}
where $\phi_*$ is calculated at $k_*=0.05$Mpc$^{-1}$, and  for each value of $f$ shown in Table \ref{table:natural_tend}.

\subsection{Hilltop quartic inflation}

\begin{figure}[th!]
\centering
\includegraphics[scale=0.5]{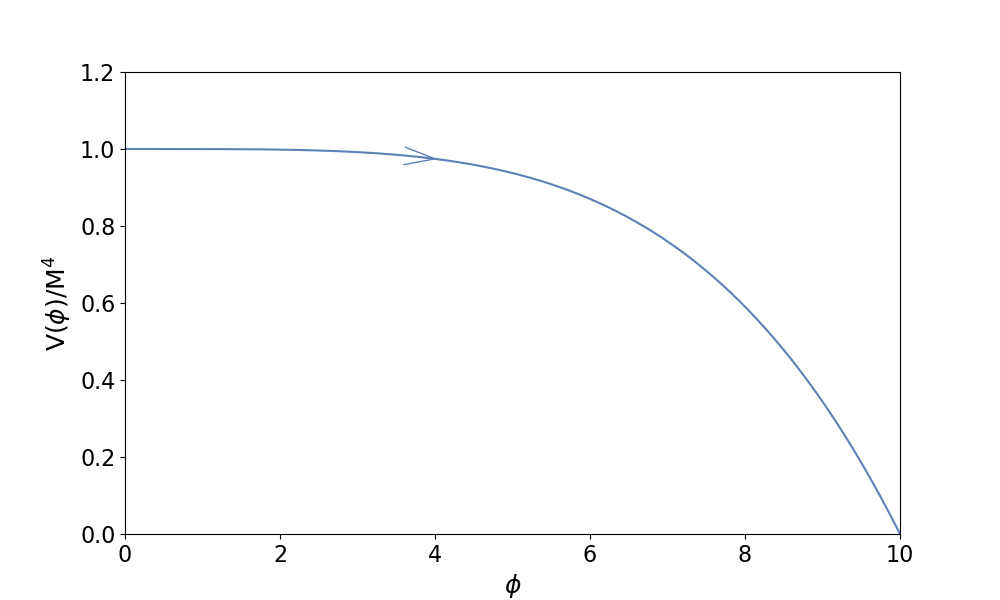}
\caption{Inflationary potential for Hilltop quartic inflation using $\mu=10$.}
\label{fig:hilltop_V}
\end{figure}

Here, we consider the Hilltop quartic inflation model that fits with the Planck $2018$ observational data \cite{akrami:2020,martin:2014,martin:2006,cook:2023,german:2021,stein:2023,hoffmann:2023,lillepalu:2023,dimopoulos:2020,kallosh:2019}. The Hilltop inflationary potential is given by \cite{boubekeur:2005},

\begin{equation}
\label{hilltop_V}
V(\phi)=M^4 \left[1-\left(\dfrac{\phi}{\mu}\right)^p\right],
\end{equation}
where $M$ must be fixed via the amplitude of the scalar power spectrum $P_\sca(k)$ given by Eq. \eqref{M},  $-2 < \log_{10} \mu < 2$ \cite{akrami:2020}, and $p=4$. The form of the potential is shown in Fig. \ref{fig:hilltop_V}. Similarly as in natural inflation the field $\phi$ rolls down from left to right, decreasing from close to the maximum of the potential towards its minimum \cite{antusch:2015}. In fact, this kind of Hilltop model has also been considered in the context of warm inflation \cite{sanchez:2008}. 

From Eqs. \eqref{epsilon} and \eqref{eta} the slow--roll parameters for this model are given by,

\begin{eqnarray}
\label{hilltop_epsilon}
\epsilon&=&\dfrac{p^2}{2\mu^2} \dfrac{\left(\frac{\phi}{\mu}\right)^{2p-2}}{\left[1-\left(\frac{\phi}{\mu}\right)^p\right]^2},\\
\label{hilltop_eta}
\eta&=&-\dfrac{p (p-1)}{\mu^2} \dfrac{\left(\frac{\phi}{\mu}\right)^{p-2}}{\left[1-\left(\frac{\phi}{\mu}\right)^p\right]}.
\end{eqnarray}

Inflation ends when $\epsilon=1$, hence we calculate the expression for $\phi_{\textnormal{end}}$,

\begin{equation}
\label{hilltop_phiend}
\left(\dfrac{\phi_{\textnormal{end}}}{\mu}\right)^p+\dfrac{p}{\sqrt{2}\mu}\left(\dfrac{\phi_{\textnormal{end}}}{\mu}\right)^{p-1}-1=0,
\end{equation}
which must be solved numerically. Then, the number of e--foldings (Eq. \eqref{N}) is,

\begin{equation}
\label{hilltop_N}
N\simeq \dfrac{\mu^2}{2p}\left\{\left(\dfrac{\phi_{\textnormal{ini}}}{\mu}\right)^2-\left(\dfrac{\phi_{\textnormal{end}}}{\mu}\right)^2 -\dfrac{2}{(2-p)}\left[\left(\dfrac{\phi_{\textnormal{ini}}}{\mu}\right)^{2-p} -\left(\dfrac{\phi_{\textnormal{end}}}{\mu}\right)^{2-p}\right]\right\}.
\end{equation}

To obtain the value of $\phi_\textnormal{ini}$, for a given value of $\mu$, we set $N\sim 50-60$. Solving the slow--roll approximation [Eqs. \eqref{Hsr} and \eqref{phisr}] the scalar field $\phi_\sr(t)$ must be computed numerically from the following expression,

\begin{eqnarray}
\nonumber
\dfrac{\phi_\textnormal{sr} }{(2-p)}&\left(\dfrac{\phi_{\textnormal{sr}}}{\mu}\right)^{1-p}& {_2}F_{1}\left[-\dfrac{1}{2},-1+\dfrac{2}{p},\dfrac{2}{p},\left(\dfrac{\phi_\mathrm{sr}}{\mu}\right)^p\right]\\
\label{hilltop_phisr}
-\dfrac{\phi_\textnormal{ini} }{(2-p)}&\left(\dfrac{\phi_{\textnormal{ini}}}{\mu}\right)^{1-p}& {_2}F_{1}\left[-\dfrac{1}{2},-1+\dfrac{2}{p},\dfrac{2}{p},\left(\dfrac{\phi_\mathrm{ini}}{\mu}\right)^p\right]-\dfrac{p M^2}{\sqrt{3}\mu} t=0,
\end{eqnarray}
where $\phi_{\textnormal{sr}}$ depends explicitly on the physical time $t$, and the parameter $M$ is fixed from Eq. \eqref{M},

\begin{equation}
\label{hilltop_M}
M^4=\dfrac{12 \delta_R p^2 \pi^2\left(\dfrac{\phi_*}{\mu}\right)^{2p}}{\phi_*^{2}\left[\left(\dfrac{\phi_*}{\mu}\right)^{p}-1\right]^3},
\end{equation}
where $\phi_*$ is calculated at the horizon crossing $k=0.05$ Mpc$^{-1}$.

We consider the special case $p=4$, hence Eq. \eqref{hilltop_phisr} becomes,

\begin{eqnarray}
\label{hilltop_phisrp4}
\nonumber
&-&\mu\dfrac{\sqrt{1-\left(\frac{\phi_\textnormal{sr}}{\mu}\right)^4}}{2\left(\frac{\phi_\textnormal{sr}}{\mu}\right)^2}+\mu \arctan\left[\dfrac{\sqrt{1-\left(\frac{\phi_\textnormal{sr}}{\mu}\right)^4}}{1+\left(\frac{\phi_\textnormal{sr}}{\mu}\right)^2}\right]\\
&+&\mu\dfrac{\sqrt{1-\left(\frac{\phi_\mathrm{ini}}{\mu}\right)^4}}{2\left(\frac{\phi_\mathrm{ini}}{\mu}\right)^2}
-\mu \arctan\left[\dfrac{\sqrt{1-\left(\frac{\phi_\mathrm{ini}}{\mu}\right)^4}}{1+\left(\frac{\phi_\mathrm{ini}}{\mu}\right)^2}\right]-\dfrac{4 M^2}{\sqrt{3}\mu} t=0,
\end{eqnarray}
which is also solved numerically.

To solved the equation of motion numerically we integrated Eqs. \eqref{H^2} and \eqref{ddotphi}, using as initial condition the numerical derivative of $\phi_\sr$ evaluated at $t=0$, which is calculated also numerically from  Eqs. \eqref{Hsr} and \eqref{phisr}. In both cases the integration was done since $t_{\ini}=0$ until $t_{\en}$ considering $p=4$. In table \ref{table:hilltop_tend} we show the value of $t_{\en}$ for each value of $\mu$ that we use in our contour plots for the Hilltop inflation model for $N=50$, and $N=60$.

\begin{table}[th!]
\begin{center}
\begin{tabular}{c | c | c }
\toprule 
N & $\mu$ & $t_\en \times 10^6$ $\left[\textnormal{M}_\Pl ^{-1}\right]$\\
\midrule 
\multirow{9}{*}{$50$}  & 7 & $8.06819$  \\
                       & 9 & $5.92022$ \\
                       & 11 & $4.80594$ \\
                       & 13 & $4.14827$ \\
                       & 15 & $3.72415$ \\
                       & 18&  $3.31895$\\
                       & 25&  $2.84391$\\
                       & 35& $2.55985$\\
                       & 95& $2.21469$\\
\midrule
\multirow{9}{*}{$60$}  & 7 & $10.24704$  \\
                       & 9 & $8.94586$ \\
                       & 11 & $7.12657$ \\
                       & 13 & $6.05813$ \\
                       & 15 & $5.37257$ \\
                       & 18&  $4.72167$\\
                       & 25&  $3.96524$\\
                       & 35& $3.51943$\\
                       & 95& $2.98903$\\
\bottomrule
\end{tabular}
\caption{Value of the time at the end of inflation $t_\en$ for each value of $\mu$ in the Hilltop inflation scenario.}
\label{table:hilltop_tend}
\end{center}
\end{table}

The scalar power spectrum $P_\sca(k)$, the scalar spectral index $n_\sca(k)$, and the tensor--to--scalar ratio $r$ are obtained from Eqs. \eqref{sr_PS}, \eqref{sr_nS}, and \eqref{sr_r}; and they are given by,

\begin{eqnarray}
\label{hilltop_PSsr}
\nonumber
P_\sca(k)&=&\dfrac{M^4 }{144 \pi^2 p^2 }\left\{-12\phi_*^2 \left[\left(\frac{\phi_*}{\mu}\right)^p-1\right]^2+p \left(\frac{\phi_*}{\mu}\right)^p\left\{4-4p-4\left(\frac{\phi_*}{\mu}\right)^p \right.  \right. \\
\nonumber
&+&\left.  5p\left(\frac{\phi_*}{\mu}\right)^p
 +12 C\left[\left(\frac{\phi_*}{\mu}\right)^p+p-1\right]\Biggl\} \right\}\left(\dfrac{\phi_*}{\mu}\right)^{-2p}\left[\left(\frac{\phi_*}{\mu}\right)^p-1\right],\\
 \\
n_\sca(k)&=& 1+p\left[2-2p-\left(2+p\right)\left(\frac{\phi_*}{\mu}\right)^p\right]\dfrac{\left(\frac{\phi_*}{\mu}\right)^p}{\phi_*^2\left[\left(\frac{\phi_*}{\mu}\right)^p-1\right]^2},\\
\label{hilltop_r}
r&=&8 p^2\dfrac{\left(\frac{\phi_*}{\mu}\right)^{2p}}{\phi_*^2\left[\left(\frac{\phi_*}{\mu}\right)^p-1\right]^2},
\end{eqnarray}
where $\phi_*$ is calculated at $k_*=0.05$Mpc$^{-1}$, and for each value of $\mu$ shown in Table \ref{table:hilltop_tend}.

\subsection{Starobinsky Inflationary Model}

\begin{figure}[th!]
\centering
\includegraphics[scale=0.5]{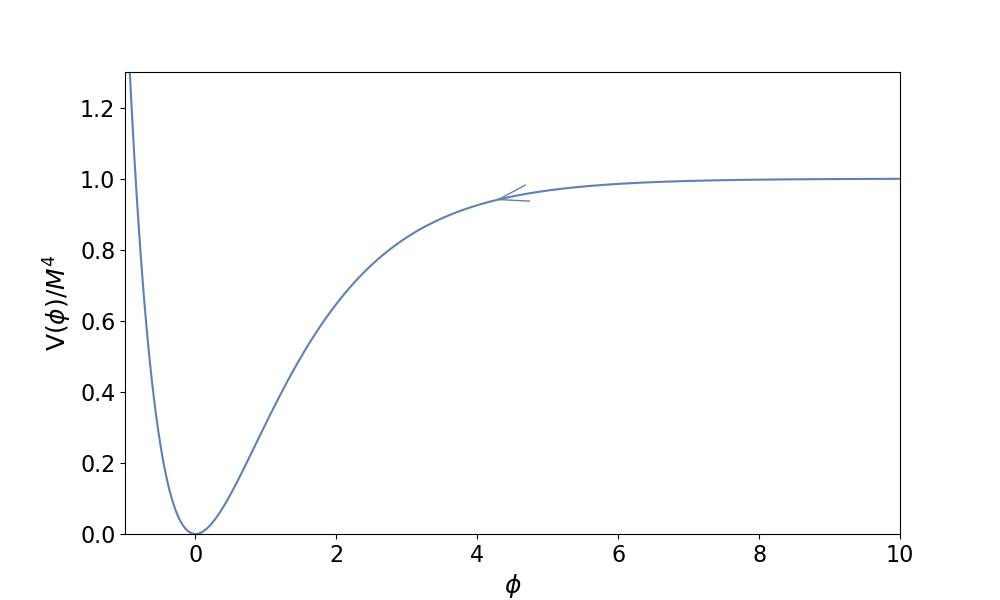}
\caption{Inflationary potential for the Starobinsky inflationary model.}
\label{fig:higgs_V}
\end{figure}

This scenario was introduced in $1980$ by A. Starobinsky \cite{starobinsky:1980} and has become the best inflationary model supported by Planck data \cite{akrami:2020}. In particular, this scheme yields a small tensor--to--scalar ratio $r$. The form of the inflationary potential is given by \cite{,martin:2014,rojas:2022,martin:2019,diValentino:2017},

\begin{equation}
\label{starobinsky_V}
V(\phi)=M^4 \left(1-e^{-\sqrt{\frac{2}{3}}\phi}\right)^2,
\end{equation}
where $M$ is a parameter to be fixed by the amplitude of the scalar power spectrum $P_\sca(k)$ given by Eq. \eqref{M}. Inflation happens from right to left and here the scalar field $\phi$ rolls slowly close to its maximum until its minimum. The behavior of the Starobinsky inflationary potential is shown in Fig. \ref{fig:higgs_V}.

The slow--roll parameters are calculated from Eqs. \eqref{epsilon} and \eqref{eta}, yielding,

\begin{eqnarray}
\label{higgs_epsilon}
\epsilon&=&\dfrac{4}{3} \dfrac{1}{\left(e^{\sqrt{\frac{2}{3}}\phi}-1\right)^2},\\
\label{higgs_eta}
\eta&=&-\dfrac{4}{3} \dfrac{\left(e^{\sqrt{\frac{2}{3}}\phi}-2\right)}{\left(e^{\sqrt{\frac{2}{3}}\phi}-1\right)^2}.
\end{eqnarray}

Once again $\epsilon=1$ indicates that inflation has ceased, we compute $\phi_\textnormal{end}$ from Eq. \eqref{higgs_epsilon}, having,

\begin{equation}
\label{higgs_phiend}
\phi_\textnormal{end}=\sqrt{\dfrac{3}{2}}\ln\left(\dfrac{2+\sqrt{3}}{\sqrt{3}}\right).
\end{equation}

Then, the number of e--folding is giving by,

\begin{equation}
\label{higgs_N}
N\simeq \dfrac{3}{4} e^{\sqrt{\frac{2}{3}}\phi_\textnormal{ini}}-\dfrac{\sqrt{3}}{{2\sqrt{2}}} \phi_\textnormal{ini}-\dfrac{3}{4} e^{\sqrt{\frac{2}{3}}\phi_\textnormal{end}}+\dfrac{\sqrt{3}}{{2\sqrt{2}}} \phi_\textnormal{end}.
\end{equation}

Fixing $N\sim 50 -60$ e--folds we compute the initial value of the scalar field $\phi_\textnormal{ini}$,

\begin{eqnarray}
\label{higgs_phiini}
\nonumber
\phi_\textnormal{ini}&=&-\sqrt{\dfrac{3}{2}} \left(1+\dfrac{2}{\sqrt{3}}\right)-2\sqrt{\dfrac{2}{3}}N+\sqrt{\dfrac{3}{2}}\ln\left(1+\dfrac{2}{\sqrt{3}}\right)\\
&-&\dfrac{2}{\sqrt{3}} W_{-1} \left[ -e^{-\left(1+\frac{2}{\sqrt{3}}\right)-\frac{4N}{3}+\ln\left(1+\frac{2}{\sqrt{3}}\right)}\right].
\end{eqnarray}

The slow--roll equations allow us to obtain $a_{\sr}(t)$ and $\phi_{\sr}(t)$, yielding,

\begin{eqnarray}
\label{higgs_asr}
\nonumber
a_\sr(t)&\simeq&\exp\left\{-\dfrac{3}{4}\left[-\dfrac{4}{3\sqrt{3}}M^2 t+\ln\left(e^{\sqrt{\frac{2}{3}}\phi_\textnormal{ini}}\right)-\ln\left(-\dfrac{4}{3\sqrt{3}}M^2 t +e^{\left(e^{\sqrt{\frac{2}{3}}\phi_\textnormal{ini}}\right)}\right)\right]\right\},\\\\
\label{higgs_phisr}
\phi_\sr(t)&\simeq&\sqrt{\dfrac{3}{2}}\ln\left(-\dfrac{4}{3\sqrt{3}}M^2 t+e^{\sqrt{\frac{2}{3}}\phi_\textnormal{ini}}\right),
\end{eqnarray}
where $M$ is computed from Eq. \eqref{M} and is equal to,

\begin{equation}
\label{starobinsky_M}
M^4=\dfrac{32 \delta_R e^{2\sqrt{\frac{2}{3}}\phi_*}\pi^2}{\left(e^{2\sqrt{\frac{2}{3}}\phi_*}-1\right)^4},
\end{equation}
where $\phi_*$ is calculated at the horizon crossing $k=0.05$ Mpc$^{-1}$.

To solved the equation of motion numerically we integrated Eqs. \eqref{H^2} and \eqref{ddotphi}, using as initial condition the numerical derivative of $\phi_\sr$ evaluated at $t=0$, which is calculated  from  Eqs. \eqref{higgs_phisr}. The integration was done since $t_{\ini}=0$ until $t_{\en}$. The initial condition for $\dot\phi_{\sr}(0)$ is given by,

\begin{equation}
\dot\phi_{\sr}(0) =-\dfrac{2}{3}\sqrt{2}M^2 e^{-\sqrt{\frac{2}{3}}\phi_\ini}. 
\end{equation}

In table \ref{table:higgs_tend} we show the value of $t_{\en}$ for $N=50$, and $N=60$. 

\begin{table}[th!]
\begin{center}
\begin{tabular}{c | c }
\toprule 
N  & $t_\en \times 10^6$ $\left[\textnormal{M}_\Pl ^{-1}\right]$\\
\midrule 
\multirow{1}{*}{$50$}   & $6.71723$  \\
\midrule
\multirow{1}{*}{$60$}   & $9.94144$  \\
\bottomrule
\end{tabular}
\caption{Value of the time at the end of inflation $t_\en$ for each value of $\mu$ in the Starobinsky inflationary model.}
\label{table:higgs_tend}
\end{center}
\end{table}

\newpage
The scalar power spectrum $P_\sca(k)$, the spectral index $n_\sca(k)$, and the tensor--to--scalar ratio $r$ are given by,

\begin{eqnarray}
\label{higgs_PSsr}
\nonumber
P_\sca(k)&=&\dfrac{M^4}{288 \pi^2}\left[-1-2\left(7+6C\right) e^{\sqrt{\frac{2}{3}}\phi_*}+9e^{2\sqrt{\frac{2}{3}}\phi_*}\right] e^{-2\sqrt{\frac{2}{3}}\phi_*}\left(e^{\sqrt{\frac{2}{3}}\phi_*}-1\right)^2,\\
\\
\label{higgs_nS}
n_\sca(k)&=&\dfrac{-5-14 e^{\sqrt{\frac{2}{3}}\phi_*}+3 e^{2\sqrt{\frac{2}{3}}\phi_*}}{3\left(\e^{\sqrt{\frac{2}{3}}\phi_*}-1\right)^2},\\
\label{higgs_r}
r&=&\dfrac{64}{3\left(\e^{\sqrt{\frac{2}{3}}\phi_*}-1\right)^2},
\end{eqnarray}
where $\phi_*$ is calculated at the horizon crossing $k_*=0.05$Mpc$^{-1}$.

\subsection{Large field inflationary models}

\begin{figure}[th!]
\centering
\includegraphics[width=\textwidth]{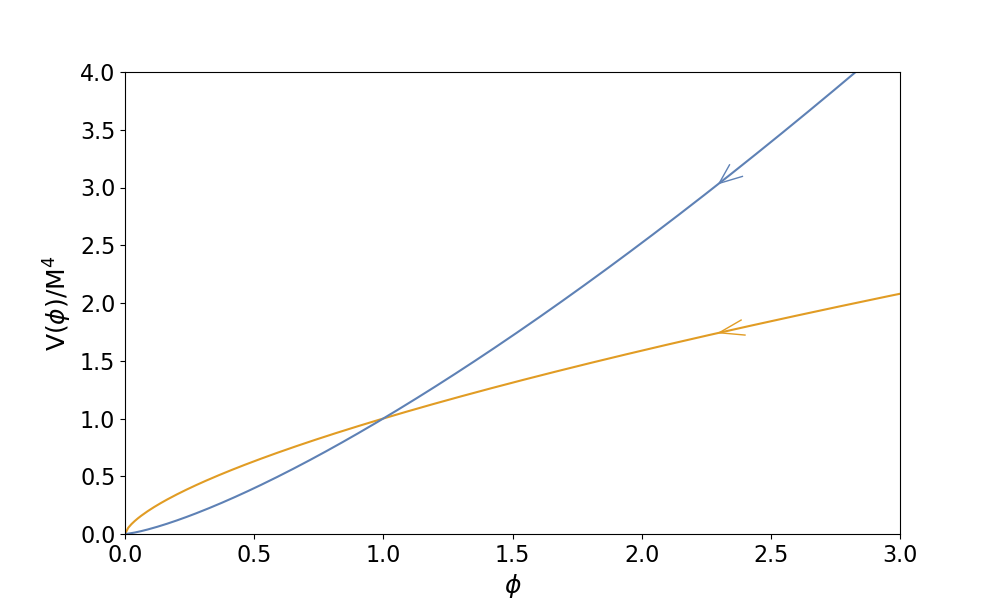}
\caption{Potential $\sfrac{V(\phi)}{M^{4}}$ versus $\phi$ for large field inflationary models. The orange line corresponds to $p=\sfrac{2}{3}$, and the blue line to $p=\sfrac{4}{3}$.}	
\label{fig:LFI}
\end{figure}

The large field inflationary models were first introduced by A. D. Linde in $1983$ \cite{martin:2006,linde:1983}. These models depends on the parameter $p$, where the potential component is given by the expression \cite{akrami:2020,martin:2019,martin:2014},

\begin{equation}
V(\phi)=M^4 \phi^p,
\end{equation}
here $M$ is fixed with respect to the amplitude of the scalar power spectrum $P_\sca(k)$ given by Eq. \eqref{M}. Models with $p>2$ are already ruled out from observations; however, instances with $p=1$ \cite{mcallister:2010} and $p=\sfrac{2}{3}$ are more favored by Planck data \cite{akrami:2020}. Here, in this work, we consider $p=\sfrac{2}{3}$ \cite{mcallister:2014,silverstein:2008} and $p=\sfrac{4}{3}$ \cite{mcallister:2014}. The forms of these potentials are shown in Fig. \ref{fig:LFI}. Similarly in the Starobinsky case, inflation occurs from right to left, and the field $\phi$ slows down near its maximum until it reaches its minimum.

The slow--roll parameters are computed from Eqs. \eqref{epsilon} and \eqref{eta}, which are given by,

\begin{eqnarray}
\epsilon&=&\dfrac{1}{2} \dfrac{p^2}{\phi^2},\\
\eta&=&\dfrac{p(p-1)}{\phi^2}.
\end{eqnarray}

Recall that inflation ends when $\epsilon=1$, so we obtain the final value of the scalar field $\phi_\textnormal{end}$,

\begin{equation}
\label{LFI_phiend}
\phi_\textnormal{end}=\dfrac{p}{\sqrt{2}}.
\end{equation}

Then, the number of e--foldings is given by,

\begin{equation}
N=\dfrac{1}{2p} \left(\phi_\textnormal{ini}^2-\phi_\textnormal{end}^2 \right).
\end{equation}

We set the value of $N\sim 50-60$ e--folds, then we obtain the initial value for the scalar field $\phi_\textnormal{ini}$, having,

\begin{equation}
\phi_\textnormal{ini}=\sqrt{ 2 p N +\dfrac{p^2}{2}}.
\end{equation}

Solving the slow--roll equations allow us to find $a_{\sr}(t)$ and $\phi_{\sr}(t)$, that is,

\begin{eqnarray}
\nonumber
\label{LFI_asr}
a_\textnormal{sr}(t)&=&\textnormal{Exp}\left\{\dfrac{1}{2p}\left\{ \left(\phi_{\textnormal{ini}}^{2-\sfrac{p}{2}}\right)^{\sfrac{4}{(4-p)}}-\left[{
\phi_{\textnormal{ini}}^{2-\sfrac{p}{2}}}+\dfrac{M^2 p (p-4)}{2\sqrt{3}}t\right]\right\}^{\sfrac{4}{(4-p)}}\right\},\\
\\
\label{LFI_phisr}
\phi_\textnormal{sr}(t)&=&\left[\phi_{\textnormal{ini}}^{2-\sfrac{p}{2}}+\dfrac{M^2 p (p-4)}{2 \sqrt{3}}t\right]^{\sfrac{4}{(4-p)}},
\end{eqnarray}
where the parameter $M$ is calculated from Eq. \eqref{M} and is given by,

\begin{equation}
\label{LFI_M}
M^4=\dfrac{12 \delta_R p^2 \pi^2}{\phi_*^{p+2}},
\end{equation}
where $\phi_*$ is calculated at the horizon crossing $k=0.05$ Mpc$^{-1}$.

To solved the equation of motion numerically we integrated Eqs. \eqref{H^2} and \eqref{ddotphi}, using as initial condition the numerical derivative of $\phi_\sr$ evaluated at $t=0$, which is calculated  from  Eq. \eqref{LFI_phisr}. The integration was done since $t_{\ini}=0$ until $t_{\en}$ considering $p=\sfrac{2}{3}$, and $p=\sfrac{4}{3}$.  The initial condition for $\dot\phi_{\sr}(0)$ is in each case given by,

\begin{eqnarray}
\dot\phi_{\sr}(0) &=& -\dfrac{2 M^2}{3 \sqrt{3} \,\phi_\ini^{\sfrac{2}{3}}}, \quad \textnormal{for} \quad p=\sfrac{2}{3},\\
\dot\phi_{\sr}(0) &=& -\dfrac{4 M^2}{3 \sqrt{3} \,\phi_\ini^{\sfrac{1}{3}}}, \quad \textnormal{for} \quad p=\sfrac{4}{3}.
\end{eqnarray}

Table \ref{table:LFI_tend} shows the value of $t_{\en}$ for each value of $p$ that we use in our contour plots for the large field inflationary model for $N=50$, and $N=60$.

\begin{table}[th!]
\begin{center}
\begin{tabular}{c  | c | c }
\toprule 
N  & $p$ & $t_\en \times 10^6$ $\left[\textnormal{M}_\Pl ^{-1}\right]$\\
\midrule 
\multirow{2}{*}{$50$}  & $\sfrac{2}{3}$ & $2.31907$  \\
                       & $\sfrac{4}{3}$  & $1.96204$ \\
\midrule
\multirow{2}{*}{$60$}  & $\sfrac{2}{3}$ & $3.09884$  \\
                       & $\sfrac{4}{3}$  & $2.77418$ \\
\bottomrule
\end{tabular}
\caption{Value of the time at the end of inflation $t_\en$ for each value of $p$ in the large field inflationary model.}
\label{table:LFI_tend}
\end{center}
\end{table}

\newpage
The scalar power spectrum $P_\sca(k)$, the scalar spectral index $n_\sca(k)$, and the tensor--to--scalar ratio $r$ are given by,

\begin{eqnarray}
\label{LFI_PSsr}
P_\sca(k)&=&\dfrac{M^4}{144  \pi^2 p^2} \left[\left(4-12 C -5 p\right)p+12\phi_*^2\right] \phi_*^p,\\
\label{LFI_nS}
n_\sca(k)&=&1-\dfrac{(2+p)p}{\phi_*^2},\\
\label{LFI_r}
r&=&\dfrac{8p^2}{\phi_*^2},
\end{eqnarray}
where $\phi_*$ is calculated at the horizon crossing $k_*=0.05$ Mpc$^{-1}$.

\section{Discussions and Results}
\label{results}

In this section, we present the results using the slow--roll approximation, and those whose coming from our numerical implementation. On the one hand, we calculate the values of $A_\sca$ and $n_\sca$ at the pivot scale $k_*=0.05$\,Mpc$^{-1}$ to draw the contour plots $(A_\sca,n_\sca)$. On the other hand, the values of $r$ and $n_\sca$ are taken at the pivot scale $k_*=0.002$\,Mpc$^{-1}$, yielding the contour plots $(r,n_\sca)$. Both calculations utilize the slow--roll approximation to establish the initial conditions. Specifically, to compute $A_\sca$, $n_\sca$, and $r$ numerically, we fit the values of $P_\sca(k)$ and $P_\ten(k)$, given by the expressions \eqref{PS_fit} and \eqref{PT_fit}. Once these numbers are obtained, we then use them to calculate $n_\sca(k)$ and $r$, via Eqs. \eqref{nS} and \eqref{r}. It's important to note that the scalar amplitudes do not exactly match the amplitude set at the beginning with the normalization of $M$. The fitting process adjusts the power-law formula for the scalar spectrum to ensure an analysis that is not biased by the selection of a specific potential. As a result, while the scalar amplitude will not remain the same, it remains close to the initial COBE normalization value of $2.1 \times 10^{-9}$, typically falling within a range from $2.1 \times 10^{-9}$ to $2.214 \times 10^{-9}$ \cite{akrami:2020}.

Also, note that in all cases the upshots of $r$ and $n_\sca$ calculated from our approach are lower than the prediction given by the slow--roll approximation. On the contrary, the numerical values of $A_\sca$ are larger than those of the slow--roll implementation. This considering the relation between scalar perturbation amplitude and tensor-scalar ratio, if it is obtained that $A_\sca$ become greater in the numerical approach than in slow roll, this would affect $r$ by definition.

In the following subsections we present the data tables and the contour plots for each model of inflation discussed in section \ref{models}. In this point it is important to say that the slow--roll approximation exhibits an error between $0.5 - 1\%$ respect to the numerical results \cite{rojas:2009,rojas:2012}.

\subsection{Natural Inflation}

This model is largely characterized by the Pseudo--Nambu--Goldstone--Boson (PNGB) decay constant $f$ \cite{adams:1993,kim:2005}. Hence, different values of the observables are computed by varying this parameter. In Table \ref{table:natural_AS} we show the values of $A_\sca$ and $n_\sca$ calculated by changing $f$, with a pivot scale $k_{*} = 0.05$ \,Mpc$^{-1}$. Then, in Table \ref{table:natural_r} we show the values of $r$ and $n_\sca$, again taking several values of $f$ but this time the pivot scale is $k=0.002$\,Mpc$^{-1}$. We present two groups: $N=50$ and $N=60$. Note that for larger values of the PNGB constant $f$, the outcomes reach the ones of the Minimal Chaotic Inflation Model ($V\propto \phi^2$) \cite{escudero:2016}.

\begin{table}[th!]
\begin{center}
\begin{tabular}{c | c | c | c | c | c}
\toprule

\multirow{2}{*}{$N$} & \multirow{2}{*}{$f$}  & \multicolumn{2}{c|}{$\ln\left(10^{10}A_\sca\right)$} & \multicolumn{2}{c}{$n_\sca$} \\ 

\cline{3-6}

  &  & SRA & NR & SRA & NR \\
  
\midrule 

\multirow{8}{*}{$50$} & 4   & 3.07185 & 3.10178 & 0.928155 & 0.927331 \\ 
                      & 5   & 3.06326 & 3.08330  & 0.942521 & 0.941449 \\ 
                      & 6   & 3.05816 & 3.07471 & 0.948208 & 0.946948 \\ 
                      & 7   & 3.05499 & 3.06894 & 0.950763 & 0.949435 \\ 
                      & 9   & 3.05145 & 3.06290  & 0.952733 & 0.951429 \\ 
                      & 12  & 3.04905 & 3.05899 & 0.953547 & 0.952135 \\ 
                      & 20  & 3.04707 & 3.05706 & 0.953903 & 0.952306 \\ 
                      & 100 & 3.04603 & 3.05563 & 0.953976 & 0.952569 \\ 

\midrule

\multirow{8}{*}{$60$} & 4   & 3.0719  & 3.09701 & 0.9325654 & 0.931436 \\ 
                      & 5   & 3.06336 & 3.07810  & 0.948956  & 0.947506 \\ 
                      & 6   & 3.05822 & 3.06930  & 0.955648  & 0.954514 \\ 
                      & 7   & 3.05501 & 3.06497 & 0.958693  & 0.957331 \\ 
                      & 9   & 3.05140  & 3.05935 & 0.961051  & 0.959796 \\ 
                      & 12  & 3.04900   & 3.05535 & 0.96202   & 0.960527 \\ 
                      & 20  & 3.04701 & 3.05339 & 0.962435  & 0.961124 \\ 
                      & 100 & 3.04584 & 3.05129 & 0.962514  & 0.961283 \\ 

\bottomrule
\end{tabular}
\caption{
Natural inflation values of the amplitude of the scalar power spectrum ($A_\sca$) and scalar spectral index ($n_\sca$) at the pivot scale $k=0.05$\,Mpc$^{-1}$ changing the parameter $f$. Both observables are computed utilizing our numerical implementation (NR) and the slow--roll approximation (SRA). We present two sets, one with $N=50$, and another one with $N=60$.}
\label{table:natural_AS}
\end{center}
\end{table}

\begin{table}[th!]
\begin{center}
\begin{tabular}{c | c | c | c | c | c}
\toprule

\multirow{2}{*}{$N$} & \multirow{2}{*}{$f$}  & \multicolumn{2}{c|}{$r$} & \multicolumn{2}{c}{$n_\sca$} \\ 

\cline{3-6}

  &  & SRA & NR & SRA & NR \\
  
\midrule 

\multirow{8}{*}{$50$} & 4   & 0.0301433  & 0.028223  & 0.929964  & 0.929189 \\ 
                      & 5   &  0.059850  & 0.0571789 & 0.945037  & 0.944014 \\ 
                      & 6   &  0.084651  & 0.0816419 & 0.951059  & 0.950258 \\ 
                      & 7   &  0.103267  & 0.100273  & 0.953775  & 0.952636 \\ 
                      & 9   &  0.127134  & 0.124041  & 0.955871  & 0.954685 \\ 
                      & 12  &  0.145286  & 0.142484  & 0.956734  & 0.955274 \\ 
                      & 20  &  0.161567  & 0.158555  & 0.957108  & 0.955798 \\ 
                      & 100 &  0.170869  & 0.168063  & 0.957183  & 0.956472 \\ 

\midrule

\multirow{8}{*}{$60$} & 4   & 0.0160243  & 0.0150583 & 0.933494 & 0.932174 \\ 
                      & 5   &  0.0381866 & 0.0366104 & 0.950453 & 0.948427 \\ 
                      & 6   &  0.059092  & 0.057291  & 0.957449 & 0.956094 \\ 
                      & 7   &  0.0757757 & 0.073799  & 0.960648 & 0.959472 \\ 
                      & 9   &  0.098969  & 0.0961345 & 0.963130  & 0.962133 \\ 
                      & 12  &  0.115622  & 0.113700    & 0.964150  & 0.963000    \\ 
                      & 20  &  0.131661  & 0.129901  & 0.964585 & 0.964108 \\ 
                      & 100 &  0.140941  & 0.139356  & 0.964665 & 0.964208 \\  

\bottomrule
\end{tabular}
\caption{
Natural inflation values of the tensor--to--scalar ratio ($r$) and scalar spectral index ($n_\sca$) at the pivot scale $k=0.002$\,Mpc$^{-1}$ changing the parameter $f$. Both observables are computed utilizing our numerical implementation (NR) and the slow--roll approximation (SRA). We present two sets, one with $N=50$, and another one with $N=60$.}
\label{table:natural_r}
\end{center}
\end{table}

\begin{figure}[th!]
\centering
\begin{subfigure}[b]{0.44\textwidth}
\centering
\includegraphics[width=\textwidth]{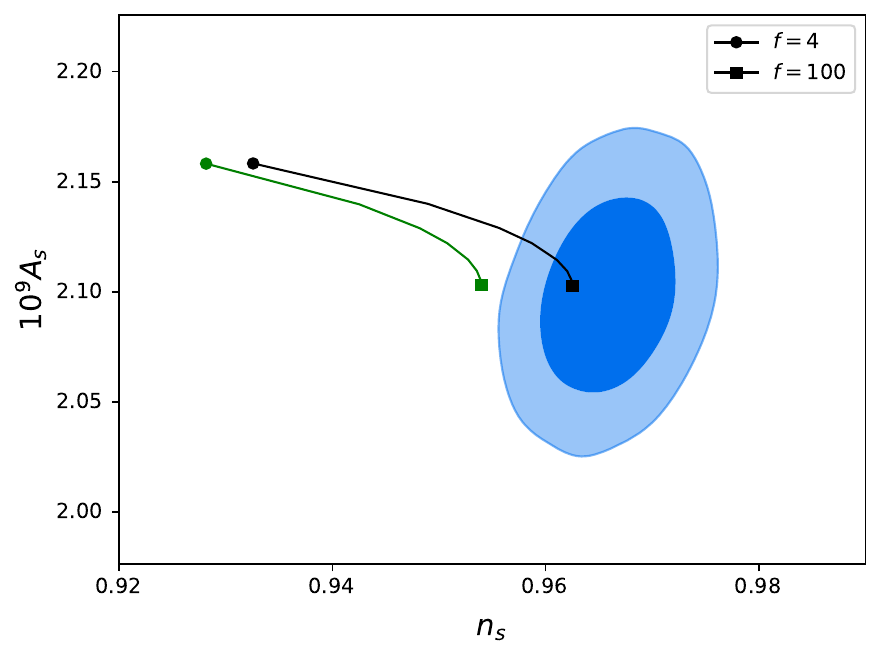}
\caption{$10^9 A_\sca$ vs $n_\sca$.}
\label{fig:natural_ASvsnS_sr}
\end{subfigure}
\hfill
\begin{subfigure}[b]{0.44\textwidth}
\centering
\includegraphics[width=\textwidth]{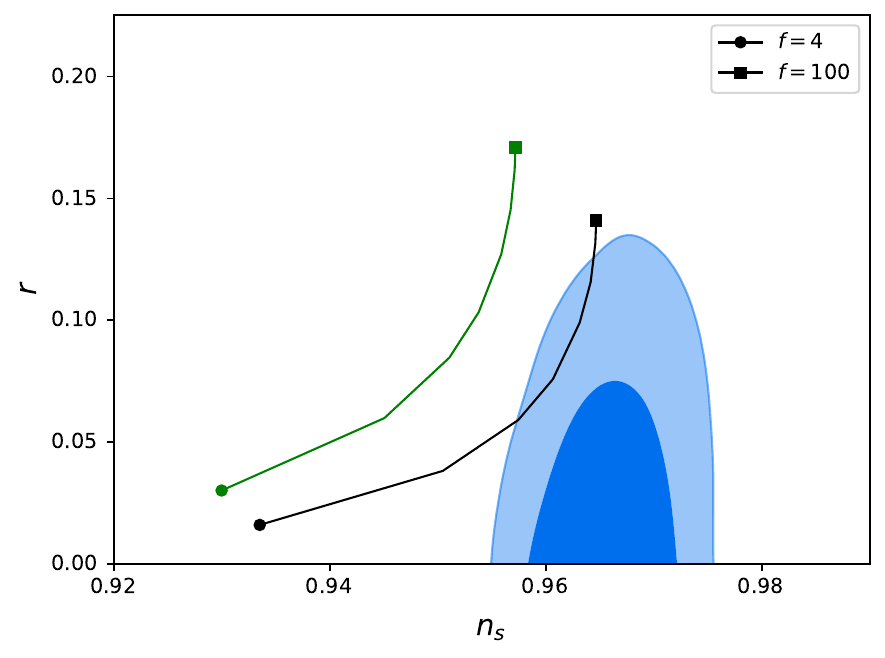}
\caption{$r$ vs $n_\sca$.}
\label{fig:natural_rvsnS_sr}
\end{subfigure}
\caption{
Contour plots $(A_\sca,n_\sca)$ and $(n_\sca,r)$ of the Natural inflation model calculated using the slow--roll approximation with $f=4,5,6,7,9,12,20,100$, where the black lines represent results with $N=60$, while the green ones describe those corresponding to $N=50$. The panel on the left--hand side (a) shows the $10^9 A_\sca$ vs $n_\sca$ evaluated at $k=0.05$\,Mpc$^{-1}$. And the right--hand side plot (b) is the $r$ vs $n_\sca$ evaluated at $k=0.002$\,Mpc$^{-1}$.}		
\label{fig:natural_contour_sr}
\end{figure}
	
\begin{figure}[th!]
\centering
\begin{subfigure}[b]{0.44\textwidth}
\centering
\includegraphics[width=\textwidth]{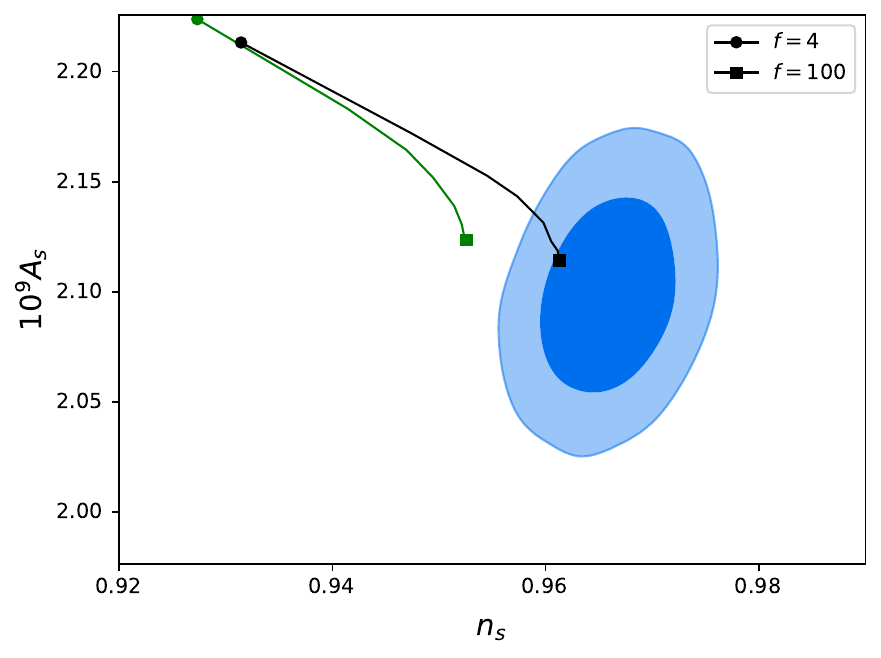}
\caption{$10^9 A_\sca$ vs $n_\sca$.}
\label{fig:natural_ASvsnS_fit}
\end{subfigure}
\hfill
\begin{subfigure}[b]{0.44\textwidth}
\centering
\includegraphics[width=\textwidth]{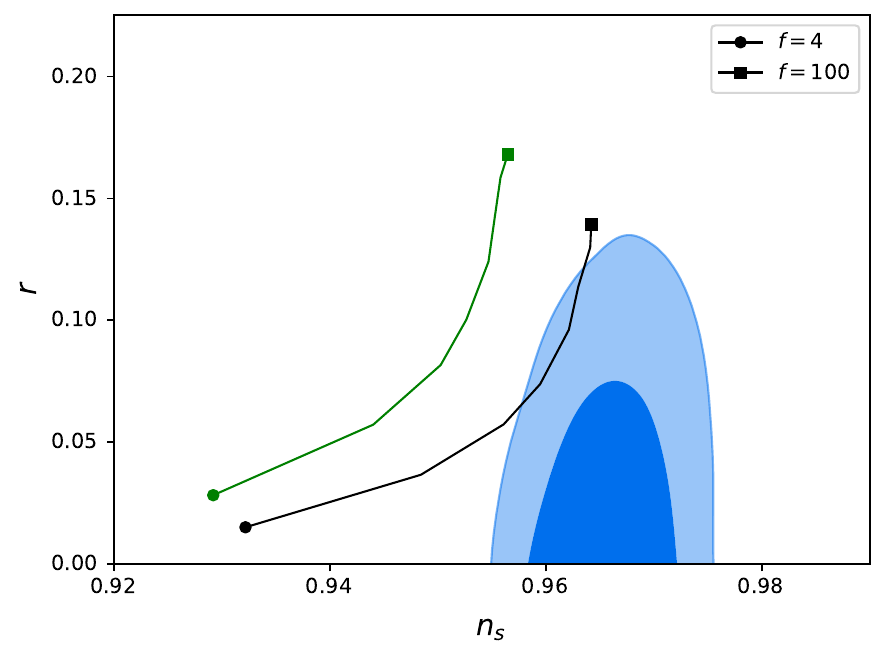}
\caption{$r$ vs $n_\sca$.}
\label{fig:natural_rvsnS_fit}
\end{subfigure}
\caption{
Contour plots $(A_\sca,n_\sca)$ and $(n_\sca,r)$ of the Natural inflation model calculated using our numerical implementation with $f=4,5,6,7,9,12,20,100$, where the black lines represent results with $N=60$, while the green ones describe those corresponding to $N=50$. The panel on the left--hand side (a) shows $10^9 A_\sca$ vs $n_\sca$ evaluated at $k=0.05$\,Mpc$^{-1}$. And the plot on the right--hand side (b) is the $r$ vs $n_\sca$ evaluated at $k=0.002$\,Mpc$^{-1}$.}	
\label{fig:natural_contour_fit}
\end{figure}

In Fig.~\ref{fig:natural_contour_sr} we present the contour plots $(A_\sca,n_\sca)$ and $(n_\sca,r)$ of the Natural inflation model calculated using the slow--roll approximation with $f=4,5,6,7,9,12,20,100$. The left--hand side panel (Fig.~\ref{fig:natural_ASvsnS_sr}) shows the $10^9 A_\sca$ vs $n_\sca$ evaluated at $k=0.05$\,Mpc$^{-1}$. And the right--hand side plot (Fig.\ref{fig:natural_rvsnS_sr}) is the $r$ vs $n_\sca$ evaluated at $k=0.002$\,Mpc$^{-1}$. On the other hand, Fig. \ref{fig:natural_contour_fit} shows the contour plots aforementioned $(A_\sca,n_\sca)$ and $(n_\sca,r)$, but these results are now obtained utilizing our numerical approach. In all plots, the black lines represent the results with $N=60$, while the green ones describe those corresponding to $N=50$.

Both slow--roll and numerical solutions exhibit a rather small range of improvement with respect to the observational data. In fact, the only outcome that still lies within this window is the case of $N = 60$, where, for instance, the black curve (in Figs.~\ref{fig:natural_ASvsnS_sr} and \ref{fig:natural_ASvsnS_fit}) slightly reaches the area of $68\%$ C.L.; and on the other hand, this same set plotted in Figs.~\ref{fig:natural_rvsnS_sr} and \ref{fig:natural_rvsnS_fit} travels across only on the $95\%$ C.L..

Finally, note that this model is relevant only for axion decay constants that exceed the Planck scale. This presents a significant theoretical challenge, since quantum gravity effects are anticipated to disrupt the shift symmetry, as well as any global symmetry, at the Planck scale. However, this is not always the case if the symmetry originates from gauge symmetry, as is possible in string theory. Nonetheless, string theory constructions typically result in a decay constant that is below the Planck scale.

\subsection{Hilltop quartic inflation}

In this subsection we present the Hilltop quartic inflation results varying the parameter $\mu$. In Table \ref{table:hilltop_AS} we show the values of $A_\sca$ and $n_\sca$, with a pivot scale $k_{*} = 0.05$ \,Mpc$^{-1}$. Then, in Table \ref{table:hilltop_r} we show the values of $r$ and $n_\sca$ but this time the pivot scale is $k=0.002$\,Mpc$^{-1}$. We display two groups: $N=50$, and $N=60$. 

\begin{table}[th!]
\begin{center}
\begin{tabular}{c | c | c | c | c | c}
\toprule

\multirow{2}{*}{$N$} & \multirow{2}{*}{$\mu$}  & \multicolumn{2}{c|}{$\ln\left(10^{10}A_\sca\right)$} & \multicolumn{2}{c}{$n_\sca$} \\ 

\cline{3-6}

  &  & SRA & NR & SRA & NR \\
  
\midrule 

\multirow{9}{*}{$50$} & 7  & 3.07052 & 3.09027  & 0.940847 & 0.940734 \\ 
                      & 9  & 3.06843 & 3.08463  & 0.945673 & 0.945371 \\ 
                      & 11 & 3.06654 & 3.08195  & 0.949481 & 0.949044 \\ 
                      & 13 & 3.06489 & 3.07888  & 0.952421 & 0.951854 \\ 
                      & 15 & 3.06350  & 3.07612  & 0.954682 & 0.954083 \\ 
                      & 18 & 3.06181 & 3.07163  & 0.957151 & 0.956450  \\ 
                      & 25 & 3.05903 & 3.06787  & 0.960424 & 0.959750  \\ 
                      & 35 & 3.05685 & 3.06545  & 0.962470  & 0.961947 \\ 
                      & 95 & 3.05321 & 3.05801  & 0.964594 & 0.963570  \\ 

\midrule

\multirow{9}{*}{$60$} & 7  & 3.06681 & 3.07965 & 0.950729 & 0.950145 \\ 
                      & 9  & 3.06519 & 3.07708 & 0.954468 & 0.953959 \\ 
                      & 11 & 3.06365 & 3.07457 & 0.957556 & 0.956850  \\ 
                      & 13 & 3.06229 & 3.07004 & 0.96003  & 0.959112 \\ 
                      & 15 & 3.06109 & 3.06864 & 0.961988 & 0.961235 \\ 
                      & 18 & 3.05960  & 3.06521 & 0.964187 & 0.963125 \\ 
                      & 25 & 3.05724 & 3.06218 & 0.967196 & 0.966246 \\ 
                      & 35 & 3.05525 & 3.0575  & 0.969133 & 0.968019 \\
                      & 95 & 3.05192 & 3.05241 & 0.971179 & 0.969734 \\ 

\bottomrule
\end{tabular}
\caption{
Hilltop quartic inflation values of the amplitude of the scalar power spectrum ($A_\sca$) and scalar spectral index ($n_\sca$) at the pivot scale $k=0.05$\,Mpc$^{-1}$ changing the parameter $\mu$. Both observables are computed utilizing our numerical implementation (NR) and the slow--roll approximation (SRA). We present two sets, one with $N=50$ and another one with $N=60$.}
\label{table:hilltop_AS}
\end{center}
\end{table}

\begin{table}[th!]
\begin{center}
\begin{tabular}{c | c | c | c | c | c}
\toprule

\multirow{2}{*}{$N$} & \multirow{2}{*}{$\mu$}  & \multicolumn{2}{c|}{$r$} & \multicolumn{2}{c}{$n_\sca$} \\ 

\cline{3-6}

  &  & SRA & NR & SRA & NR \\
  
\midrule 

\multirow{9}{*}{$50$} & 7  & 0.00348786 & 0.00331328 & 0.944542 & 0.944323 \\ 
                      & 9  & 0.00675914 & 0.00645695 & 0.948953 & 0.948583 \\ 
                      & 11 & 0.0106721  & 0.0102239  & 0.952493 & 0.952214 \\ 
                      & 13 & 0.0148423  & 0.0142607  & 0.955261 & 0.954877 \\ 
                      & 15 & 0.0189986  & 0.0183029  & 0.957412 & 0.956919 \\ 
                      & 18 & 0.0248786  & 0.0241033  & 0.959785 & 0.959142 \\ 
                      & 25 & 0.0362140   & 0.0352731  & 0.962960  & 0.963110  \\ 
                      & 35 & 0.0473473  & 0.0462421  & 0.964966 & 0.964971 \\ 
                      & 95 & 0.0703695  & 0.0691776  & 0.967064 & 0.966904 \\ 

\midrule

\multirow{9}{*}{$60$} & 7  & 0.00213601 & 0.00204658 & 0.953265 & 0.952087 \\ 
                      & 9  & 0.00430359 & 0.0041428  & 0.956743 & 0.955900   \\ 
                      & 11 & 0.00703314 & 0.0067919  & 0.959637 & 0.959287 \\ 
                      & 13 & 0.0100659  & 0.00975431 & 0.961980  & 0.961175 \\ 
                      & 15 & 0.0131925  & 0.0128254  & 0.963851 & 0.963906 \\ 
                      & 18 & 0.0177647  & 0.0173165  & 0.965969 & 0.965293 \\ 
                      & 25 & 0.0269608  & 0.0263789  & 0.968899 & 0.968298 \\ 
                      & 35 & 0.0363702  & 0.0357350   & 0.970804 & 0.969874 \\
                      & 95 & 0.0566272  & 0.0558734  & 0.972827 & 0.971642 \\ 

\bottomrule
\end{tabular}
\caption{
Hilltop quartic inflation values of the tensor--to--scalar ratio ($r$) and scalar spectral index ($n_\sca$) at the pivot scale $k=0.002$\,Mpc$^{-1}$ changing the parameter $\mu$. Both observables are computed utilizing our numerical implementation (NR) and the slow--roll approximation (SRA). We present two sets, one with $N=50$, and another one with $N=60$.}
\label{table:hilltop_r}
\end{center}
\end{table}

\begin{figure}[th!]
\centering
\begin{subfigure}[b]{0.44\textwidth}
\centering
\includegraphics[width=\textwidth]{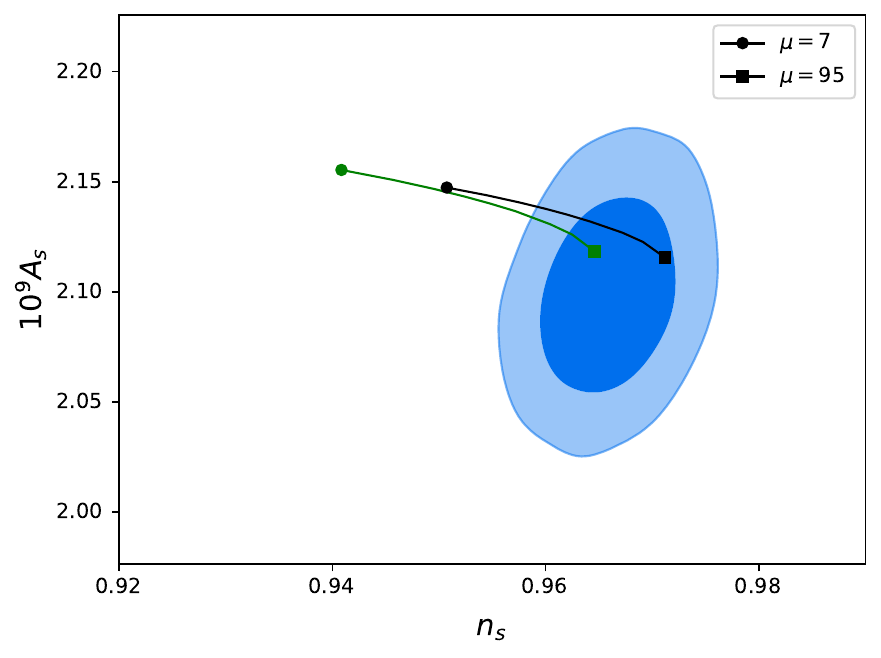}
\caption{$10^9 A_\sca$ vs $n_\sca$.}
\label{fig:hilltop_ASvsnS_sr}
\end{subfigure}
\hfill
\begin{subfigure}[b]{0.44\textwidth}
\centering
\includegraphics[width=\textwidth]{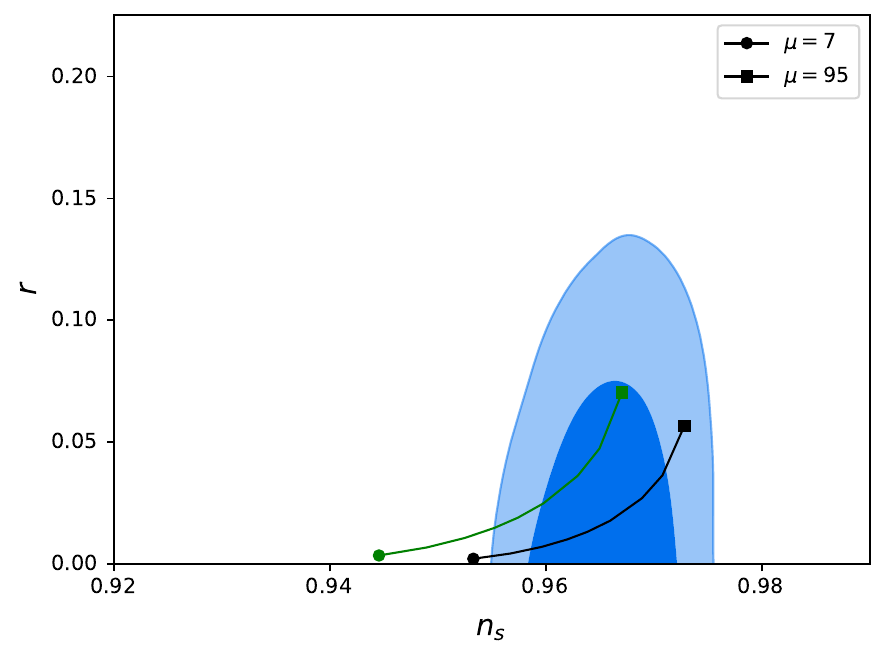}
\caption{$r$ vs $n_\sca$.}
\label{fig:hilltop_rvsnS_sr}
\end{subfigure}
\caption{
Contour plots $(A_\sca,n_\sca)$ and $(n_\sca,r)$ of the Hilltop quartic inflation model calculated using the slow--roll approximation with $\mu=7,9,11,13,15,18,25,35,95$, where the black lines represent results with $N=60$, while the green ones describe those corresponding to $N=50$. The left--hand side panel (a) shows $10^9 A_\sca$ vs $n_\sca$ evaluated at $k=0.05$\,Mpc$^{-1}$. And the plot on the right--hand side (b) is the $r$ vs $n_\sca$ evaluated at $k=0.002$\,Mpc$^{-1}$.}	
\label{fig:hilltop_contour_sr}
\end{figure}

\begin{figure}[th!]
\centering
\begin{subfigure}[b]{0.44\textwidth}
\centering
\includegraphics[width=\textwidth]{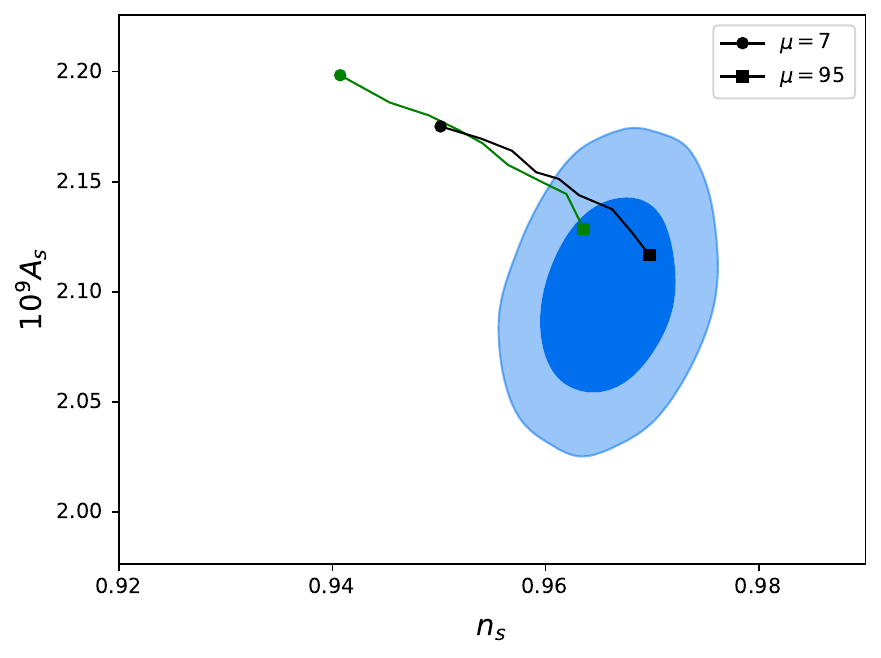}
\caption{$10^9 A_\sca$ vs $n_\sca$.}
\label{fig:hilltop_ASvsnS_fit}
\end{subfigure}
\hfill
\begin{subfigure}[b]{0.44\textwidth}
\centering
\includegraphics[width=\textwidth]{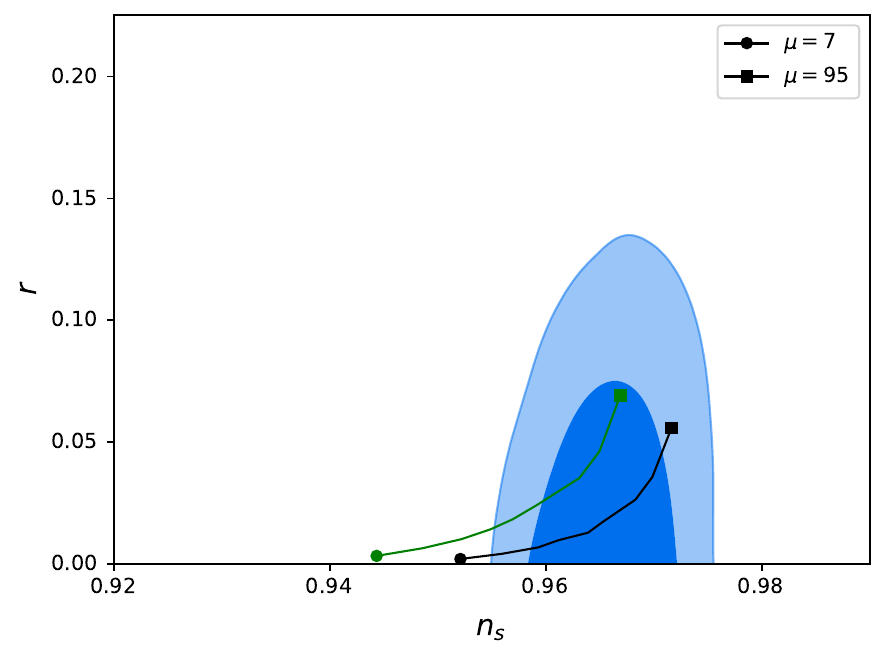}
\caption{$r$ vs $n_\sca$.}
\label{fig:hilltop_rvsnS_fit}
\end{subfigure}
\caption{
Contour plots $(A_\sca,n_\sca)$ and $(n_\sca,r)$ of the Hilltop quartic inflation model calculated using our numerical implementation with $\mu=7,9,11,13,15,18,25,35,95$, where the black lines represent results with $N=60$, while the green ones describe those corresponding to $N=50$. The left--hand side panel (a) shows $10^9 A_\sca$ vs $n_\sca$ evaluated at $k=0.05$\,Mpc$^{-1}$. And the plot on the right--hand side (b) is the $r$ vs $n_\sca$ evaluated at $k=0.002$\,Mpc$^{-1}$.}	
\label{fig:hilltop_contour_fit}
\end{figure}

Furthermore, Fig.~\ref{fig:hilltop_contour_sr} shows the contour plots $(A_\sca,n_\sca)$ and $(n_\sca,r)$ of the Hilltop quartic inflationary model calculated using the slow--roll approximation with $\mu=7,9,11,13,15,18,25,35,95$. The left--hand side panel (Fig.~\ref{fig:hilltop_ASvsnS_sr}) shows the $10^9 A_\sca$ vs $n_\sca$ evaluated at $k=0.05$\,Mpc$^{-1}$. And the right--hand side plot (Fig.\ref{fig:hilltop_rvsnS_sr}) is the $r$ vs $n_\sca$ evaluated at $k=0.002$\,Mpc$^{-1}$. On the other hand, in Fig. \ref{fig:hilltop_contour_fit} we present the contour plots aforementioned $(A_\sca,n_\sca)$ and $(n_\sca,r)$ but now these results are obtained utilizing our numerical approach. In all plots, the black lines represent the results with $N=60$, while the green ones describe those corresponding to $N=50$. Note that in both sets of results there is an important range of parameter $\mu$ that are well inside the observational area. Therefore, this potential is highly favored by the Planck 2018 data \cite{akrami:2020}.

\subsection{Starobinsky inflationary model}

The Starobinsky scenario is remarkably simple, hence only two instance are shown: one for $N = 50$ and another one for $N =60$. In Table \ref{table:starobinsky_AS} we show the values of $A_\sca$ and $n_\sca$ with a pivot scale $k_{*} = 0.05$\,Mpc$^{-1}$. Then, in Table \ref{table:starobinsky_r} we show the values of $r$ and $n_\sca$ but this time the pivot scale is $k=0.002$\,Mpc$^{-1}$. 

\begin{table}[th!]
\begin{center}
\begin{tabular}{c | c | c | c | c }
\toprule

\multirow{2}{*}{$N$}  & \multicolumn{2}{c|}{$\ln\left(10^{10}A_\sca\right)$} & \multicolumn{2}{c}{$n_\sca$} \\ 

\cline{2-5}

  & SRA & NR & SRA & NR \\
  
\midrule 

$50$ & 3.06550 & 3.07514 & 0.953663 & 0.953222 \\ 

\midrule

$60$ & 3.06204 & 3.06776 & 0.962336 & 0.961705 \\ 

\bottomrule
\end{tabular}
\caption{
Starobinsky inflation values of the amplitude of the scalar power spectrum ($A_\sca$) and scalar spectral index ($n_\sca$) at the pivot scale $k=0.05$\,Mpc$^{-1}$. Both observables are computed utilizing our numerical implementation (NR) and the slow--roll approximation (SRA). We present two sets, one with $N=50$, and another one with $N=60$.}
\label{table:starobinsky_AS}
\end{center}
\end{table}

\begin{table}[th!]
\begin{center}
\begin{tabular}{c | c | c | c | c }
\toprule

\multirow{2}{*}{$N$}  & \multicolumn{2}{c|}{$r$} & \multicolumn{2}{c}{$n_\sca$} \\ 

\cline{2-5}

  & SRA & NR & SRA & NR \\
  
\midrule 

$50$ & 0.00523324 & 0.00504663 & 0.956926 & 0.955919 \\ 

\midrule

$60$ & 0.00403100 & 0.00349071 & 0.964515 & 0.963774 \\ 

\bottomrule
\end{tabular}
\caption{
Starobinsky inflation values of the tensor--to--scalar ratio ($r$) and scalar spectral index ($n_\sca$) at the pivot scale $k=0.002$\,Mpc$^{-1}$. Both observables are computed utilizing our numerical implementation (NR) and the slow--roll approximation (SRA). We present two sets, one with $N=50$ and another one with $N=60$.}
\label{table:starobinsky_r}
\end{center}
\end{table}

\begin{figure}[th!]
\centering
\begin{subfigure}[b]{0.44\textwidth}
\centering
\includegraphics[width=\textwidth]{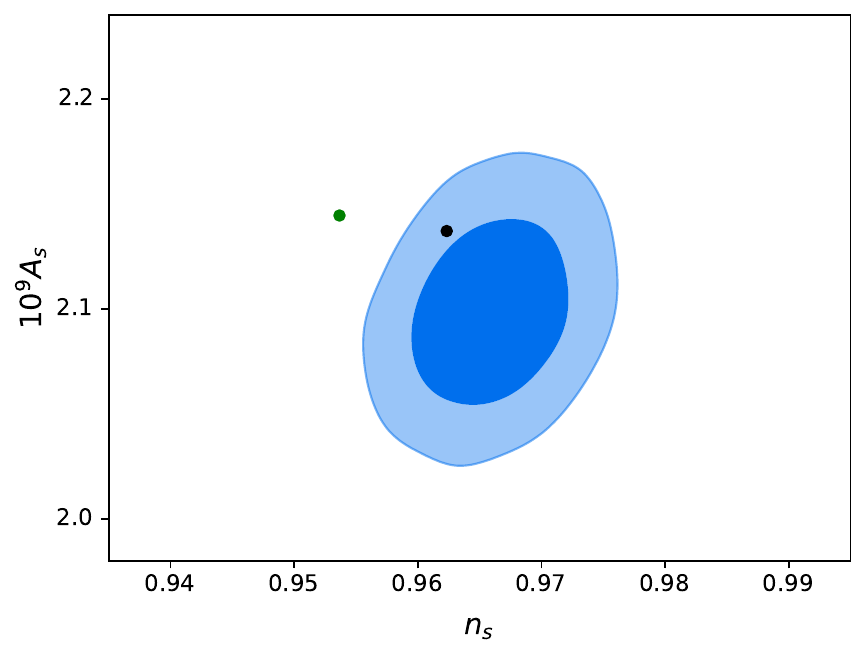}
\caption{$10^9 A_\sca$ vs $n_\sca$.}
\label{fig:higgs_ASvsnS_sr}
\end{subfigure}
\hfill
\begin{subfigure}[b]{0.44\textwidth}
\centering
\includegraphics[width=\textwidth]{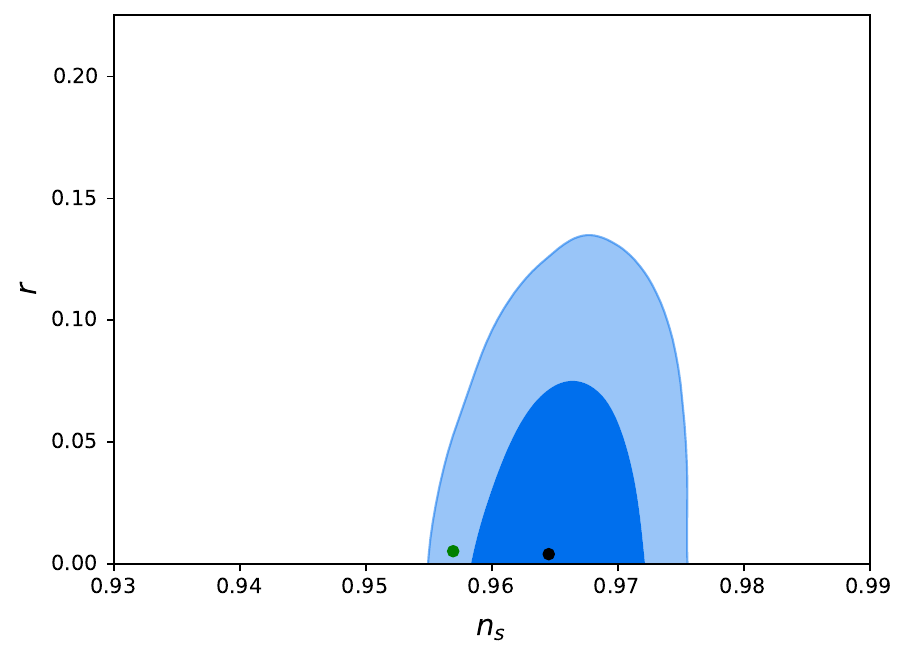}
\caption{$r$ vs $n_\sca$.}
\label{fig:higgs_rvsnS_sr}
\end{subfigure}
\caption{
Contour plots $(A_\sca,n_\sca)$ and $(n_\sca,r)$ of the Starobinsky inflationary model calculated using the slow--roll approximation, where the black dots represent the results with $N=60$, while the green ones describe those corresponding to $N=50$. The left--hand side panel (a) shows $10^9 A_\sca$ vs $n_\sca$ evaluated at $k=0.05$\,Mpc$^{-1}$. And the plot on the right--hand side (b) is the $r$ vs $n_\sca$ evaluated at $k=0.002$\,Mpc$^{-1}$.}	
\label{fig:higgs_contour_sr}
\end{figure}

\begin{figure}[th!]
\centering
\begin{subfigure}[b]{0.44\textwidth}
\centering
\includegraphics[width=\textwidth]{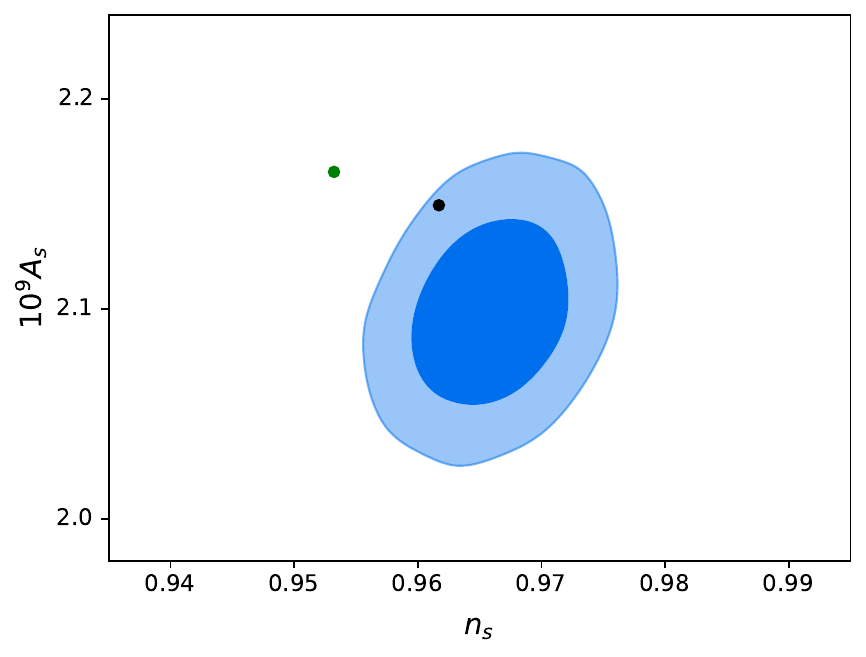}
\caption{$10^9 A_\sca$ vs $n_\sca$.}
\label{fig:higgs_ASvsnS_fit}
\end{subfigure}
\hfill
\begin{subfigure}[b]{0.44\textwidth}
\centering
\includegraphics[width=\textwidth]{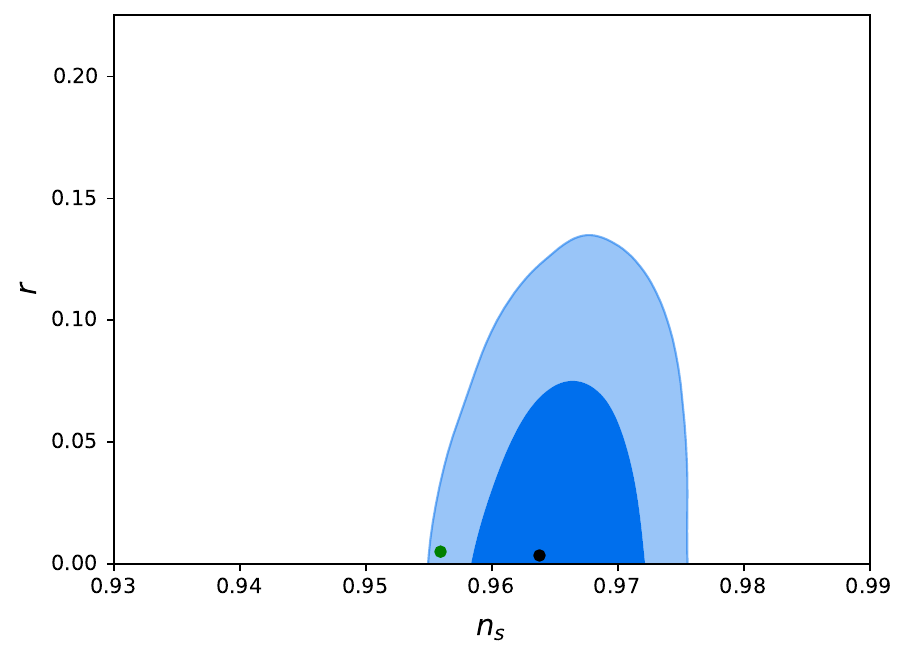}
\caption{$r$ vs $n_\sca$.}
\label{fig:higgs_rvsnS_fit}
\end{subfigure}
\caption{
Contour plots $(A_\sca,n_\sca)$ and $(n_\sca,r)$ of the Starobinsky inflationary model calculated using our numerical implementation, where the black dots represent the results with $N=60$, while the green ones describe those corresponding to $N=50$. The left--hand side panel (a) shows $10^9 A_\sca$ vs $n_\sca$ evaluated at $k=0.05$\,Mpc$^{-1}$. And the plot on the right--hand side (b) is the $r$ vs $n_\sca$ evaluated at $k=0.002$\,Mpc$^{-1}$.}	
\end{figure}

Furthermore, Fig.~\ref{fig:higgs_contour_sr} shows the contour plots $(A_\sca,n_\sca)$ and $(n_\sca,r)$ of the Starobinsky inflationary model calculated using the slow--roll approximation. The left--hand side panel (Fig.~\ref{fig:higgs_ASvsnS_sr}) shows $10^9 A_\sca$ vs $n_\sca$ evaluated at $k=0.05$\,Mpc$^{-1}$. And the right--hand side plot (Fig.\ref{fig:higgs_rvsnS_sr}) is the $r$ vs $n_\sca$ evaluated at $k=0.002$\,Mpc$^{-1}$. On the other hand, in Fig. \ref{fig:hilltop_contour_fit} we present the aforementioned contour plots $(A_\sca,n_\sca)$ and $(n_\sca,r)$ but now these results are obtained utilizing our numerical approach. In all plots, the black dots represent the results with $N=60$, while the green ones describe those corresponding to $N=50$. 

The numerical and slow--roll upshots are very similar; however, the numerical $r$ is visibly lower than the slow--roll one. Additionally, this time only the black points ($N = 60$) are well within the area of both observables. 

\subsection{Large field inflation models}

The Large field  inflationary models are also a simple scenario due to its unique parameter $p$. Hence, we present only two results with $p = \sfrac{2}{3}$ and $p = \sfrac{4}{3}$. In Table \ref{table:LFI_AS} we show the values of $A_\sca$ and $n_\sca$ with a pivot scale $k_{*} = 0.05$ \,Mpc$^{-1}$. Then, in Table \ref{table:LFI_r}  we show the values of $r$ and $n_\sca$ but this time the pivot scale is $k=0.002$\,Mpc$^{-1}$. 

\begin{table}[th!]
\begin{center}
\begin{tabular}{c | c | c | c | c | c}
\toprule

\multirow{2}{*}{$N$} & \multirow{2}{*}{$p$}  & \multicolumn{2}{c|}{$\ln\left(10^{10}A_\sca\right)$} & \multicolumn{2}{c}{$n_\sca$} \\ 

\cline{3-6}

  &  & SRA & NR & SRA & NR \\
  
\midrule 

\multirow{2}{*}{$50$} & $\sfrac{2}{3}$ & 3.05286 &  3.05574 & 0.968689 & 0.967558 \\ 
                      & $\sfrac{4}{3}$ & 3.04937  & 3.05718 & 0.961322 & 0.960497 \\ 

\midrule

\multirow{2}{*}{$60$} & $\sfrac{2}{3}$ & 3.05141 & 3.05172 & 0.974593 & 0.973154 \\ 
                      & $\sfrac{4}{3}$ & 3.04861 & 3.05272 & 0.968546 & 0.967468 \\ 

\bottomrule
\end{tabular}
\caption{
Large field power--law potentials inflation values of the amplitude of the scalar power spectrum ($A_\sca$) and scalar spectral index ($n_\sca$) at the pivot scale $k=0.05$\,Mpc$^{-1}$ with $p=\sfrac{2}{3}, \sfrac{4}{3}$. Both observables are computed utilizing our numerical implementation (NR) and the slow--roll approximation (SRA). We present two sets, one with $N=50$ and another one with $N=60$.}
\label{table:LFI_AS}
\end{center}
\end{table}

\begin{table}[th!]
\begin{center}
\begin{tabular}{c | c | c | c | c | c}
\toprule

\multirow{2}{*}{$N$} & \multirow{2}{*}{$p$}  & \multicolumn{2}{c|}{$r$} & \multicolumn{2}{c}{$n_\sca$} \\ 

\cline{3-6}

  &  & SRA & NR & SRA & NR \\
  
\midrule 

\multirow{2}{*}{$50$} & $\sfrac{2}{3}$ & 0.0582048 & 0.0572722 & 0.970898 & 0.969751 \\ 
                      & $\sfrac{4}{3}$ & 0.115105  & 0.113198  & 0.96403  & 0.963526 \\ 

\midrule

\multirow{2}{*}{$60$} & $\sfrac{2}{3}$ & 0.0478681 & 0.0472768 & 0.976066 & 0.974887 \\ 
                      & $\sfrac{4}{3}$ & 0.0948539 & 0.0936424 & 0.970358 & 0.969395 \\  

\bottomrule
\end{tabular}
\caption{
Large field power--law potentials inflation values of the tensor--to--scalar ratio ($r$) and scalar spectral index ($n_\sca$) at the pivot scale $k=0.002$\,Mpc$^{-1}$ with $p=\sfrac{2}{3}, \sfrac{4}{3}$. Both observables are computed utilizing our numerical implementation (NR) and the slow--roll approximation (SRA). We present two sets, one with $N=50$, and another one with $N=60$.}
\label{table:LFI_r}
\end{center}
\end{table}

\begin{figure}[th!]
\centering
\begin{subfigure}[b]{0.44\textwidth}
\centering
\includegraphics[width=\textwidth]{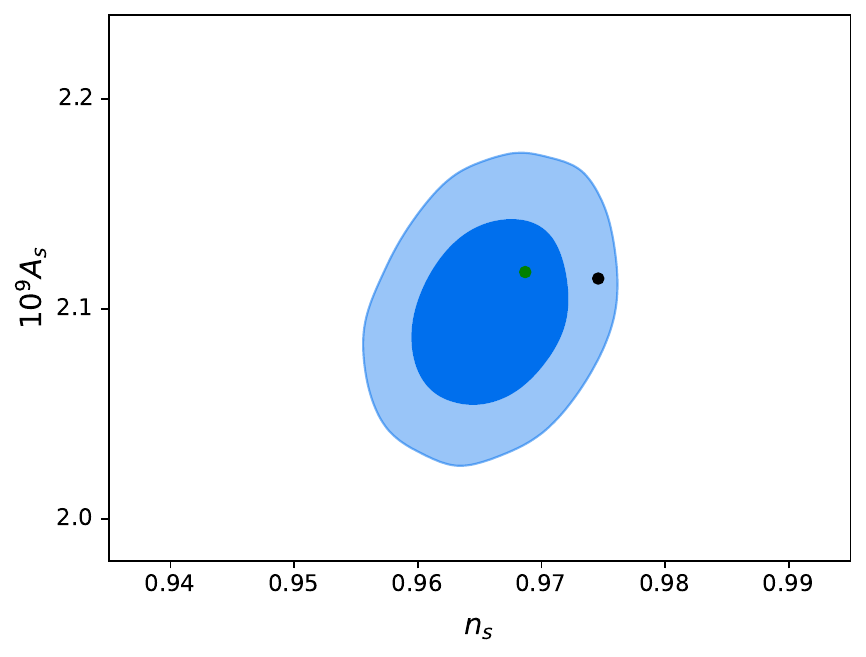}
\caption{$10^9 A_\sca$ vs $n_\sca$ for $p=\sfrac{2}{3}$.}
\label{fig:LFI_ASvsnS_p=2.3_sr}
\end{subfigure}
\hfill
\begin{subfigure}[b]{0.44\textwidth}
\centering
\includegraphics[width=\textwidth]{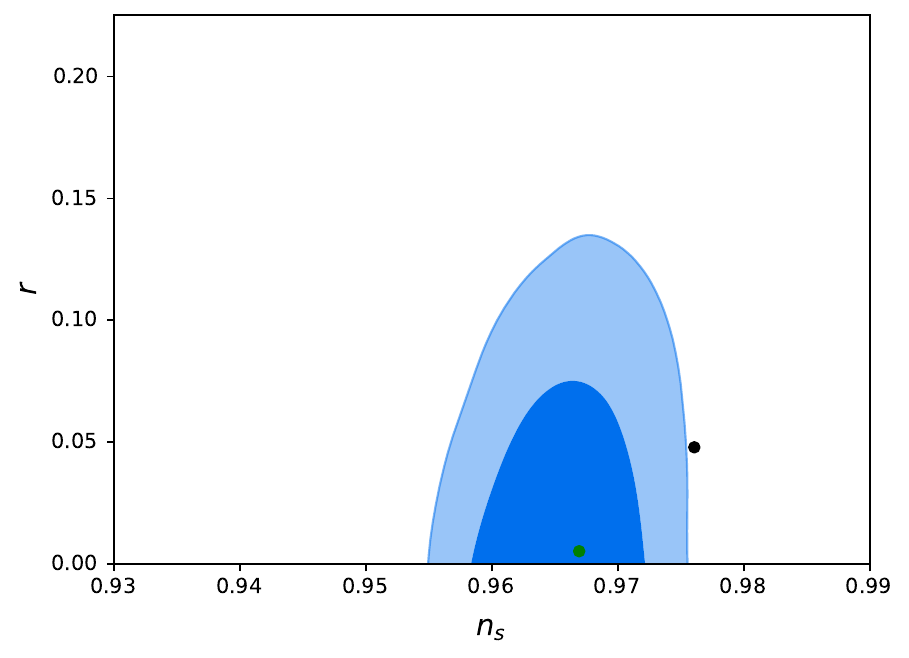}
\caption{$r$ vs $n_\sca$ for $p=\sfrac{2}{3}$.}
\label{fig:LFI_rvsnS_p=2.3_sr}
\end{subfigure}
\begin{subfigure}[b]{0.44\textwidth}
\centering
\includegraphics[width=\textwidth]{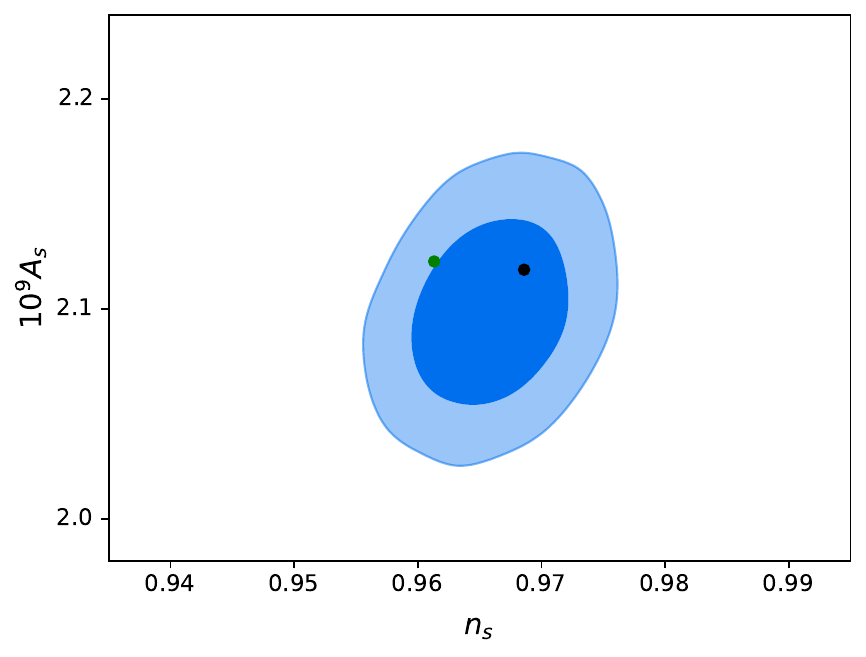}
\caption{$10^9 A_\sca$ vs $n_\sca$ for $p=\sfrac{4}{3}$.}
\label{fig:LFI_ASvsnS_p=4.3_sr}
\end{subfigure}
\hfill
\begin{subfigure}[b]{0.44\textwidth}
\centering
\includegraphics[width=\textwidth]{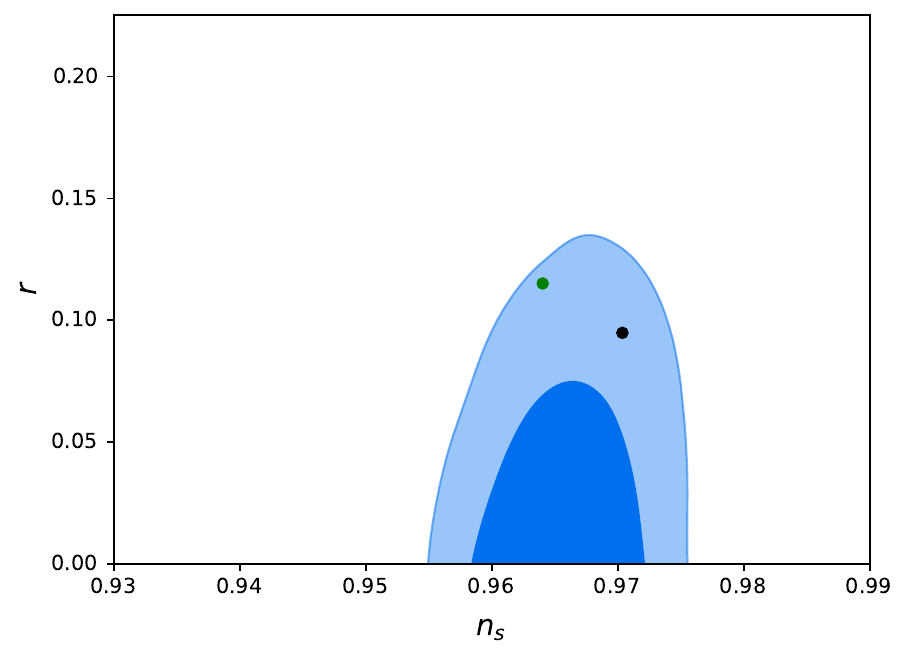}
\caption{$r$ vs $n_\sca$ for $p=\sfrac{4}{3}$.}
\label{fig:LFI_rvsnS_p=4.3_sr}
\end{subfigure}
\caption{
Contour plots $(A_\sca,n_\sca)$ and $(n_\sca,r)$ of the Large field power--law potentials model calculated using the slow--roll approximation, where the black dots represent the results with $N=60$, while the green ones describe those corresponding to $N=50$. The left panels show the $10^9 A_\sca$ vs $n_\sca$ evaluated at $k=0.05$\,Mpc$^{-1}$: (a) with $p=\sfrac{2}{3}$ and (c) $p=\sfrac{4}{3}$. And the right plots are the $r$ vs $n_\sca$ evaluated at $k=0.002$\,Mpc$^{-1}$: (b) with $p=\sfrac{2}{3}$, and (d) $p=\sfrac{4}{3}$.}		
\label{fig:LFI_contours_p=2.3_sr}
\end{figure}	

\begin{figure}[th!]
\centering
\begin{subfigure}[b]{0.45\textwidth}
\centering
\includegraphics[width=\textwidth]{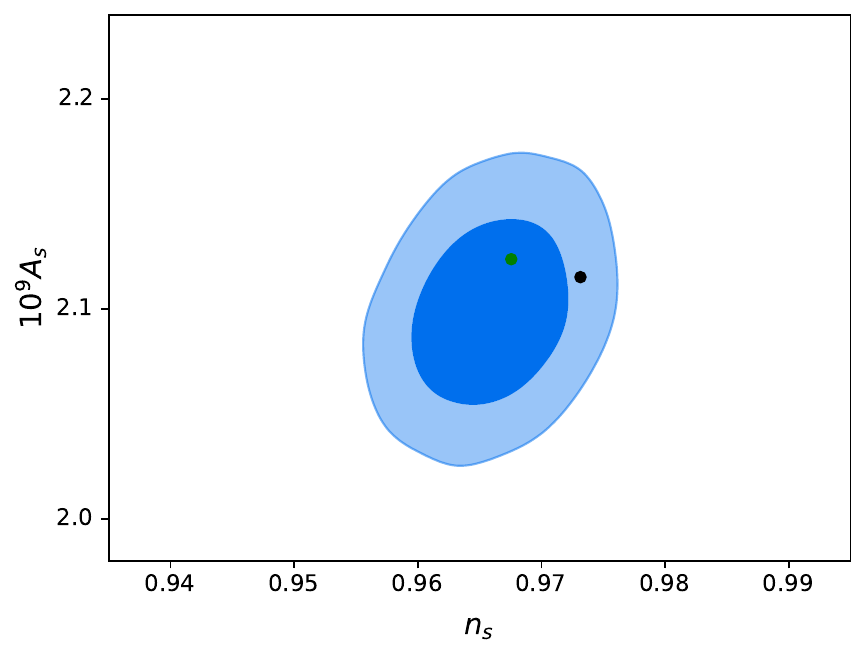}
\caption{$10^9 A_\sca$ vs $n_\sca$ for $p=\sfrac{2}{3}$.}
\label{fig:LFI_ASvsnS_p=2.3_fit}
\end{subfigure}
\hfill
\begin{subfigure}[b]{0.45\textwidth}
\centering
\includegraphics[width=\textwidth]{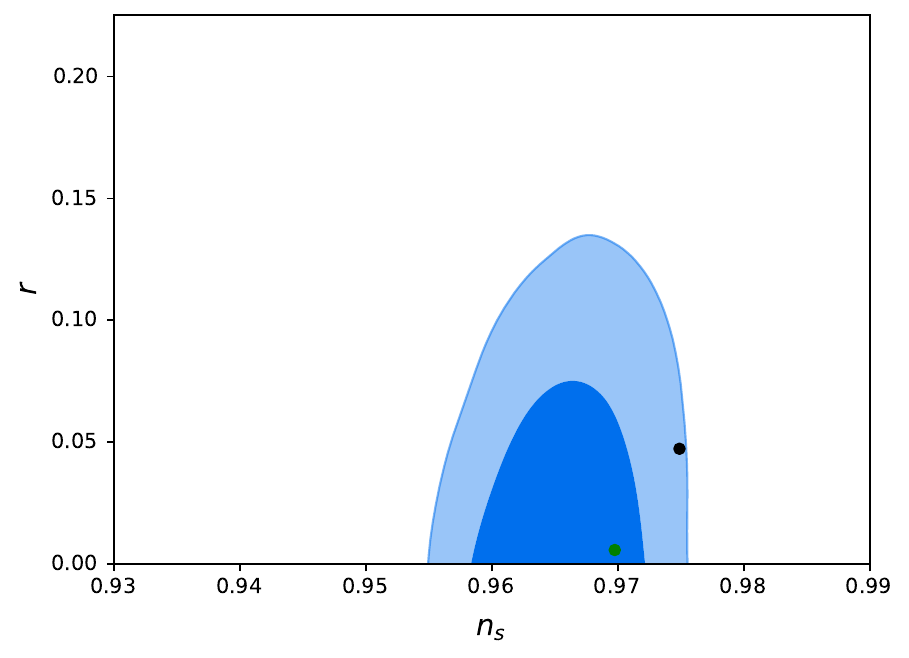}
\caption{$r$ vs $n_\sca$ for $p=\sfrac{2}{3}$.}
\label{fig:LFI_rvsnS_p=2.3_fit}
\end{subfigure}
\begin{subfigure}[b]{0.45\textwidth}
\centering
\includegraphics[width=\textwidth]{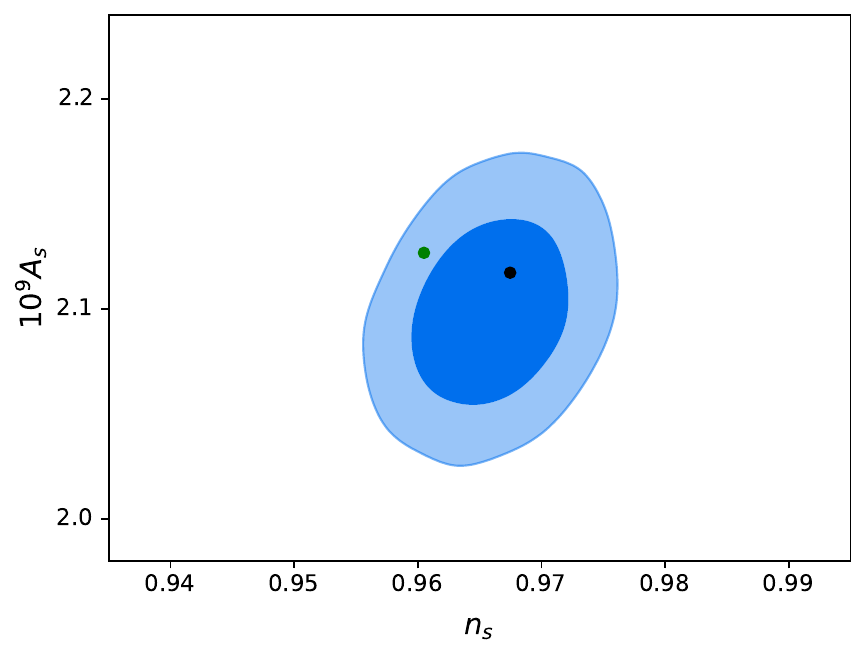}
\caption{$10^9 A_\sca$ vs $n_\sca$ for $p=\sfrac{4}{3}$.}
\label{fig:LFI_ASvsnS_p=4.3_fit}
\end{subfigure}
\hfill
\begin{subfigure}[b]{0.45\textwidth}
\centering
\includegraphics[width=\textwidth]{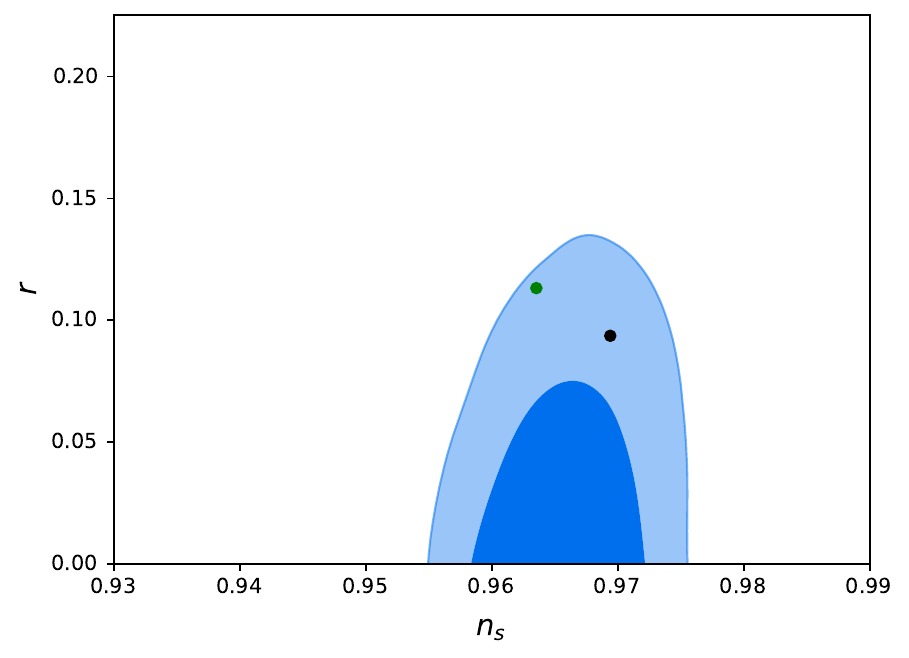}
\caption{$r$ vs $n_\sca$ for $p=\sfrac{4}{3}$.}
\label{fig:LFI_rvsnS_p=4.3_fit}
\end{subfigure}
\caption{
Contour plots $(A_\sca,n_\sca)$ and $(n_\sca,r)$ of the Large field power-law potentials model calculated using our numerical implementation, where the black dots represent the results with $N=60$, while the green ones describe those corresponding to $N=50$. The left panels show the $10^9 A_\sca$ vs $n_\sca$ evaluated at $k=0.05$\,Mpc$^{-1}$: (a) with $p=\sfrac{2}{3}$ and (c) $p=\sfrac{4}{3}$. And the right plots are the $r$ versus $n_\sca$ evaluated in $k=0.002$\,Mpc$^{-1}$: (b) with $p=\sfrac{2}{3}$, and (d) $p=\sfrac{4}{3}$.}		
\label{fig:LFI_contours_p=2.3_fit}
\end{figure}

Furthermore, Fig.~\ref{fig:LFI_contours_p=2.3_sr} shows the contour plots $(A_\sca,n_\sca)$ and $(n_\sca,r)$ of the Large field power--law potentials model calculated using the slow--roll approximation. The left panels (Figs.~\ref{fig:LFI_ASvsnS_p=2.3_sr} and~\ref{fig:LFI_ASvsnS_p=4.3_sr}) show the $10^9 A_\sca$ vs $n_\sca$ evaluated at $k=0.05$\,Mpc$^{-1}$. And the right plots (Figs.~\ref{fig:LFI_rvsnS_p=2.3_sr} and~\ref{fig:LFI_rvsnS_p=4.3_sr} ) is the $r$ vs $n_\sca$ evaluated at $k=0.002$\,Mpc$^{-1}$. On the other hand, in Fig. \ref{fig:LFI_contours_p=2.3_fit} we present the aforementioned contour plots $(A_\sca,n_\sca)$ and $(n_\sca,r)$ but now these results are obtained utilizing our numerical approach. In all plots, the black dots represent the results with $N=60$, while the green ones describe those corresponding to $N=50$. 

The analysis of two distinct cases is as follows. For $p=\sfrac{2}{3}$ at $N=50$ both the numerical and slow--roll outcomes lie within the observation window (at $68\%$ C.L.). In contrast, at $N=60$ only the numerical upshot is contained at the $95\%$ C.L.; hence our method improves this prediction. On the other hand, for example, $p=\sfrac{4}{3}$ all points sit inside $95\%$ C.L. or $68\%$ C.L., and the single contrast emerges at $N=50$ since the numerical solution shifts the point from the $1\sigma$'s to the $2\sigma$'s zones. 

\section*{Acknowledgements}
\label{acknoledgements}

One of the authors, C. Rojas, wants to thank Gabriel Germ\'an, and Jessica Cook for useful discussions. R.H.J. is supported by CONAHCYT Estancias Posdoctorales por M\'{e}xico, Modalidad 1: Estancia Posdoctoral Acad\'{e}mica and by SNI-CONAHCYT. 

\section{Conclusions}\label{conclusions}

In this study, we investigated four widely recognized inflationary scenarios, as presented by the most recent Planck observations: Natural inflation, Hilltop quartic inflation, the Starobinsky inflationary model, and Large field power--law potentials, denoted by $V(\phi)\sim \phi^{p}$, where we considered specific values such as $p=\sfrac{2}{3}$ and $\sfrac{4}{3}$. 

Our analysis mainly consisted of a comparison between the slow--roll approximation and our numerical implementation; where the latter yields better results. On the one hand, we calculate the values of $A_\sca$ and $n_\sca$ at the pivot scale $k_*=0.05$\,Mpc$^{-1}$ to draw the contour plots $(A_\sca,n_\sca)$. On the other hand, the values of $r$ and $n_\sca$ are taken at the pivot scale $k_*=0.002$\,Mpc$^{-1}$, yielding the contours plots $(r,n_\sca)$. Both calculations utilize the slow--roll approximation to set the initial conditions.

In Natural inflation, both slow--roll and numerical solutions demonstrated a relatively limited degree of enhancement concerning the observational data. In fact, the only outcome that remained consistent within this range was the case of $N = 60$.

The Hilltop quartic inflation model is characterized by $\eta$, and in both sets of results, there is an important range of this parameter that is well inside the observational area. Thus, this potential is strongly supported by the Planck 2018 data~\cite{akrami:2020}.

Although the Starobinsky scenario is remarkably simple, the case of $N = 60$ is well within the area of all observables. Moreover, both the numerical and slow--roll upshots are very similar; however, the numerical $r$ is visibly lower than the slow--roll estimation.

Finally, the Large field power--law potentials analysis of two distinct cases resulted as follows. In the scenario where $p=\sfrac{2}{3}$ at $N=50$, both the numerical and slow--roll outcomes remained within the observation window (at $68\%$ C.L.). However, when considering $N=60$, only the numerical upshot fell within the $95\%$ C.L., highlighting an enhancement in our method's predictive capability. In contrast, in the case of $p=\sfrac{4}{3}$, all data points were placed within the $95\%$ C.L. or $68\%$ C.L., except for a single deviation occurring at $N=50$ where the numerical solution shifted the point from the $1\sigma$ to the $2\sigma$ zones.

In general, our study demonstrated that the numerical solution improved the precision of these models with respect to the contour plot $r$ vs. $n_\sca$. This was evidenced by consistently lower values of $r$ in each model compared to the values derived from the slow--roll approximation. 

In conclusion, we expect this thoroughly task to provide a well--described route on the study of inflationary physics and its corresponding observational signature and its contrast with the available most current relevant data.  


\bibliographystyle{unsrt}

\begin{thebibliography}{10}
	
	\bibitem{Albrecht:1982wi}
	{A. Albrecht and P. J. Steinhardt}.
	\newblock {Cosmology for Grand Unified Theories with Radiatively Induced
		Symmetry Breaking}.
	\newblock {\em Phys. Rev. Lett.}, 48:1220, 1982.
	
	\bibitem{guth:1981}
	A.~H. Guth.
	\newblock {Inflationary universe: A possible solution to the horizon and
		flatness problems}.
	\newblock {\em Phys. Rev. D}, 23:347, 1981.
	
	\bibitem{Linde:1981mu}
	A.~D. Linde.
	\newblock {A New Inflationary Universe Scenario: A Possible Solution of the
		Horizon, Flatness, Homogeneity, Isotropy and Primordial Monopole Problems}.
	\newblock {\em Phys. Lett. B}, 108:389, 1982.
	
	\bibitem{Starobinsky:1980te}
	A.~A. Starobinsky.
	\newblock {{A New Type of Isotropic Cosmological Models Without Singularity}}.
	\newblock {\em Phys. Lett. B}, 91:99, 1980.
	
	\bibitem{Sato:1980yn}
	K.~Sato.
	\newblock {First Order Phase Transition of a Vacuum and Expansion of the
		Universe}.
	\newblock {\em Mon. Not. Roy. Astron. Soc.}, 195:467, 1981.
	
	\bibitem{weinberg:2008}
	S.~Weinberg.
	\newblock {\em Cosmology}.
	\newblock OUP Oxford, 2008.
	
	\bibitem{vazquez:2020}
	{J. A. Vazquez, L. E. Padilla, and T. Matos}.
	\newblock {Inflationary cosmology: from theory to observations}.
	\newblock {\em Rev. Mex. Fis. E}, 17:73, 2020.
	
	\bibitem{preskill:1979}
	{J. Preskill}.
	\newblock {Cosmological Production of Superheavy Magnetic Monopoles}.
	\newblock {\em Phys. Rev. Lett.}, 43:1365, 1979.
	
	\bibitem{odintsov:2023}
	{S. D. Odintsov, V. K. Oikonomou, I. Giannakoudi, F. P. Fronimos, and E. C.
		Lymperiadou}.
	\newblock {Recent Advances on Inflation}.
	\newblock {\em Symmetry}, 15:1701, 2023.
	
	\bibitem{martin:2008}
	{J. Martin}.
	\newblock {Inflationary perturbations: The cosmological Schwinger effect}.
	\newblock {\em Lecture Notes in Physics}, 738:193, 2008.
	
	\bibitem{akrami:2020}
	Y.~Akrami \textit{et al.}
	\newblock {Planck 2018 results. X. Constraints on inflation}.
	\newblock {\em Astron. Astrophys.}, 641:A10, 2020.
	
	\bibitem{habib:2005b}
	{S. Habib and A. Heinen and K. Heitmann and G. Jungman}.
	\newblock {Inflationary Perturbations and Precision Cosmology}.
	\newblock {\em Phys. Rev. D}, 71:043518, 2005.
	
	\bibitem{giare:2023c}
	{W. Giar\`e, M. De Angelis, C. van de Bruck, and E. Di Valentino }.
	\newblock {Tracking the Multifield Dynamics with Cosmological Data: A Monte
		Carlo approach}.
	\newblock {\em JCAP}, 12:014, 2023.
	
	\bibitem{german:2023}
	{G. Germ\'an, J. C. Hidalgo, and L. E. Padilla}.
	\newblock {Inflationary models constrained by reheating}.
	\newblock {\em arXiv:1404.6704}, 2023.
	
	\bibitem{martin:2014}
	{J. Martin, C. Ringeval, and V. Vennin}.
	\newblock {Encyclopaedia Inflationaris}.
	\newblock {\em Phys. Dark Univ.}, 5--6:75, 2014.
	
	\bibitem{martin:2006}
	{J. Martin, and C. Ringeval}.
	\newblock {Inflation after WMAP3: confronting the slow--roll and exact power
		spectra with CMB data}.
	\newblock {\em JCAP}, 24:009, 2006.
	
	\bibitem{freese:1990}
	{K. Freese, J. A. Frieman, and A. V. Olinto}.
	\newblock {Natural inflation with pseudo Nambu--Goldstone bosons}.
	\newblock {\em Phys. Rev. Lett.}, 65:3233, 1990.
	
	\bibitem{adams:1993}
	{F. C. Adams, J. R. Bond, K. Freese, J. A. Fireman, and A. V. Olinto}.
	\newblock {Natural inflation: Particle physics models, power--law spectra for
		large--scale structure, and constraints from the Cosmic Background Explorer}.
	\newblock {\em Phys. Rev. D}, 47:426, 1993.
	
	\bibitem{boubekeur:2005}
	{L. Boubekeur and D. H. Lyth}.
	\newblock {Hilltop inflation}.
	\newblock {\em JCAP}, 07:010, 2005.
	
	\bibitem{starobinsky:1980}
	A.~A. Starobinsky.
	\newblock {A new type of isotropic cosmological models without singularity}.
	\newblock {\em Phys. Lett. B}, 91:99, 1980.
	
	\bibitem{linde:1983}
	{A. D. Linde}.
	\newblock {Chaotic Inflation}.
	\newblock {\em Phys. Rev. D}, 129B:177, 1983.
	
	\bibitem{liddle:2000}
	A.~R. Liddle and D.~H. Lyth.
	\newblock {\em {Cosmological inflation and large--scale structure}}.
	\newblock Cambridge University Press, 2000.
	
	\bibitem{copeland:1993}
	{E. J. Copeland, E. W . Kolb, A. R. Liddle, and J. E. Lidsey}.
	\newblock {Observing the Inflaton Potential}.
	\newblock {\em Phys. Rev. Letts}, 71, 1993.
	
	\bibitem{adshead:2011}
	{P. Adshead, R. Easther, J. Pritchard, and A. Loeb}.
	\newblock { Inflation and the scale dependent spectral index: prospects and
		strategies}.
	\newblock {\em JCAP}, 02:021, 2011.
	
	\bibitem{ragavendra:2023}
	{H. V. Ragavendra, and L. Sriramkumar}.
	\newblock {Observational Imprints of Enhanced Scalar Power on Small Scales in
		Ultra Slow Roll Inflation and Associated Non--Gaussianities }.
	\newblock {\em Galaxies}, 11, 2023.
	
	\bibitem{giare:2023b}
	{W. Giar\`e, S. Pan, E. Di Valentino, W. Yang, J. de Haro, and A. Melchiorri}.
	\newblock {Inflationary Potential as seen from Different Angles: Model
		Compatibility from Multiple CMB Missions}.
	\newblock {\em arXiv:2305.15378v1}, 2023.
	
	\bibitem{vazquez:2013}
	{J. A. Vazquez, M. Bridges, Y--Z. Ma, and M. P. Hobson}.
	\newblock {Constraints on the tensor--to--scalar ratio for non--power--law
		models}.
	\newblock {\em JCAP}, 08:001, 2013.
	
	\bibitem{das:2023}
	{S. Das and R. O. Ramos}.
	\newblock {Running and Running of the Running of the Scalar Spectral Index in
		Warm Inflation}.
	\newblock {\em Universe}, 9:76, 2023.
	
	\bibitem{finelli:2018}
	{F. Finelli {\textit et al.}}
	\newblock {Exploring cosmic origins with CORE: Inflation}.
	\newblock {\em JCAP}, 04:016, 2018.
	
	\bibitem{freese:2015}
	{K. Freese, and W. H. Kinney}.
	\newblock {Natural inflation: consistency with cosmic microwave background
		observations of Planck and BICEP2}.
	\newblock {\em JCAP}, 03:044, 2015.
	
	\bibitem{stein:2022}
	{N. K. Stein, and W. H. Kinney}.
	\newblock {Natural inflation after Plack 2018}.
	\newblock {\em JCAP}, 01:022, 2022.
	
	\bibitem{nina:2022}
	{N. K. Stein, and W. H. Kinney}.
	\newblock {Natural inflation after Plack 2018}.
	\newblock {\em JCAP}, 01:022, 2022.
	
	\bibitem{montefalcone:2023}
	{G. Montefalcone, V. Aragam, L. Visinelli, and K. Freese}.
	\newblock {Observational constrainsts on warm natural inflation}.
	\newblock {\em JCAP}, 03:002, 2023.
	
	\bibitem{cook:2023}
	{J. L. Cook}.
	\newblock {Primordial Black Hole Production in Natural and Hilltop Inflation}.
	\newblock {\em JCAP}, 07:031, 2023.
	
	\bibitem{german:2021}
	{G. Germ\'an}.
	\newblock {Quartic hilltop inflation revisited}.
	\newblock {\em JCAP}, 02:034, 2021.
	
	\bibitem{stein:2023}
	{N. K. Stein, and W. H. Kinney}.
	\newblock {Simple single--field inflation models with arbitrarily small
		tensor/scalar ratio}.
	\newblock {\em JCAP}, 03:027, 2022.
	
	\bibitem{hoffmann:2023}
	{J. Hoffmann and D. Sloan}.
	\newblock {Regularization of Single Field Inflation Models}.
	\newblock {\em Phys. Rev. D}, 2023.
	
	\bibitem{lillepalu:2023}
	{H. G. Lillepalu and A. Racioppi}.
	\newblock {Generalized Hilltop Inflation}.
	\newblock {\em EPJ Plus}, 138:894, 2023.
	
	\bibitem{dimopoulos:2020}
	{R. Kallosh and A. Linde}.
	\newblock {An analytic treatment of quartic hilltop inflation}.
	\newblock {\em Phys. Lett. B}, 809:135688, 2020.
	
	\bibitem{kallosh:2019}
	{R. Kallosh and A. Linde}.
	\newblock {On hilltop and brane inflation after Planck}.
	\newblock {\em JCAP}, 09:030, 2019.
	
	\bibitem{antusch:2015}
	{S. Antusch, D. Nolde, and S. Orani }.
	\newblock {Hill crossing during preheating after hilltop inflation}.
	\newblock {\em JCAP}, 06:009, 2015.
	
	\bibitem{sanchez:2008}
	{J. C. Bueno S\'anchez, M. Bestero--Gil, A. Berera, and K. Dimopoulos}.
	\newblock {Warm hilltop inflation}.
	\newblock {\em Phys. Rev. D}, 77:123527, 2008.
	
	\bibitem{rojas:2022}
	{C. Rojas}.
	\newblock {Study of scalar and tensor power spectra in the generalized
		Starobinsky inflationary model using semiclassical methods}.
	\newblock {\em Astroparticle Physics}, 143:102745, 2022.
	
	\bibitem{martin:2019}
	J.~Martin.
	\newblock {Cosmic Inflation: Trick or Treat?}
	\newblock {\em arXiv:1902.05286}, 2019.
	
	\bibitem{diValentino:2017}
	{E. Di Valentino and L. Mersini--Houghton}.
	\newblock {Testing predictions of the quantum landscape multiverse 1: the
		Starobinsky inflationary potential}.
	\newblock {\em JCAP}, 2, 2017.
	
	\bibitem{mcallister:2010}
	{L. McAllister, E. Silverstein, and A. Westphal}.
	\newblock {Gravity waves and linear inflation from axion monodromy}.
	\newblock {\em Phys. Rev. D}, 82:046003, 2010.
	
	\bibitem{mcallister:2014}
	{L. McAllister, E. Silverstein, A. Westphal, and T. Wrase}.
	\newblock {The powers of monodromy}.
	\newblock {\em JHEP}, 09:123, 2014.
	
	\bibitem{silverstein:2008}
	{E. Silverstein and A. Westphal}.
	\newblock {Monodromy in the CMB: Gravity waves and string inflation}.
	\newblock {\em Phys. Rev. D}, 78:106003, 2008.
	
	\bibitem{rojas:2009}
	{C. Rojas and V. M. Villalba}.
	\newblock {Computation of inflationary cosmological perturbations in chaotic
		inflationary scenarios using the phase--integral method}.
	\newblock {\em Phys. Rev. D}, 79:103502, 2009.
	
	\bibitem{rojas:2012}
	{C. Rojas and V. M. Villalba}.
	\newblock {Computation of the power spectrum in chaotic
		$\frac{1}{4}\lambda\phi^4$ inflation}.
	\newblock {\em JCAP}, 003:1, 2012.
	
	\bibitem{kim:2005}
	{J. E. Kim, H. P. Nilles, and M. Peloso}.
	\newblock {Completing natural inflation}.
	\newblock {\em JCAP}, 005, 2005.
	
	\bibitem{escudero:2016}
	{M. Escudero, H. Ramírez, L. Boubekeur, E. Giusarma, and O. Mena}.
	\newblock {The present and future of the most favoured inflationary models
		after Planck 2015.}
	\newblock {\em JCAP}, 020, 2016.
	
\end{thebibliography}


\end{document}